\documentclass[12pt]{article}
\newcommand{\nc}{\newcommand}
\nc{\renc}{\renewcommand}

\nc{\half}{{\textstyle{1\over2}}}
\nc{\etal}{\mbox{\it et al. }}
\nc{\ie}{{\it i.e. }}
\nc{\eg}{{\it e.g. }}

\renc{\thefootnote}{\arabic{footnote}}
\nc{\capt}[1]{{\bf Figure.} {\small\sl #1}}

\nc{\eqs}[2]{\mbox{Eqs.~(\ref{#1},\,\ref{#2})}}
\nc{\eq}[1]{\mbox{Eq.~(\ref{#1})}}

\nc{\figs}[2]{\mbox{Figs.~(\ref{#1},\,\ref{#2})}}
\nc{\fig}[1]{\mbox{Fig.~\ref{#1}}}

\nc{\tag}[1]{\label{#1} \marginpar{{\footnotesize #1}}}
\nc{\mtag}[1]{\label{#1} \mbox{\marginpar{{\footnotesize #1}}}}
\renc{\baselinestretch}{1.5}
\newlength{\overeqskip}
\newlength{\undereqskip}
\nc{\be}[1]{\begin{equation} \mbox{$\label{#1}$}}
\nc{\bea}[1]{\begin{eqnarray} \mbox{$\label{#1}$}}
\nc{\Section}[2]{\section{#2}\label{#1}}
\nc{\Bibitem}[1]{\bibitem{#1}}
\nc{\Label}[1]{\label{#1}}

\nc{\eea}{\vspace{\undereqskip}\end{eqnarray}}
\nc{\ee}{\vspace{\undereqskip}\end{equation}}
\nc{\bdm}{\begin{displaymath}}
\nc{\edm}{\end{displaymath}}
\nc{\dpsty}{\displaystyle}
\nc{\bc}{\begin{center}}
\nc{\ec}{\end{center}}
\nc{\ba}{\begin{array}}
\nc{\ea}{\end{array}}
\nc{\bab}{\begin{abstract}}
\nc{\eab}{\end{abstract}}
\nc{\btab}{\begin{tabular}}
\nc{\etab}{\end{tabular}}
\nc{\bit}{\begin{itemize}}
\nc{\eit}{\end{itemize}}
\nc{\ben}{\begin{enumerate}}
\nc{\een}{\end{enumerate}}
\nc{\bfig}{\begin{figure}}
\nc{\efig}{\end{figure}}
\nc{\arreq}{&\!=\!&}
\nc{\arrmi}{&\!-\!&}
\nc{\arrpl}{&\!+\!&}
\nc{\arrap}{&\!\!\!\approx\!\!\!&}
\nc{\align}{\!\!\!\!\!\!\!\!&&}

\def\lsim{\; \raise0.3ex\hbox{$<$\kern-0.75em
      \raise-1.1ex\hbox{$\sim$}}\; }
\def\gsim{\; \raise0.3ex\hbox{$>$\kern-0.75em
      \raise-1.1ex\hbox{$\sim$}}\; }
\nc{\DOT}{\hspace{-0.08in}{\bf .}\hspace{0.1in}}
\nc{\Laada}{\hbox {$\sqcap$ \kern -1em $\sqcup$}}
\nc\loota{{\scriptstyle\sqcap\kern-0.55em\hbox{$\scriptstyle\sqcup$}}}
\nc\Loota{{\sqcap\kern-0.65em\hbox{$\sqcup$}}}
\nc\laada{\Loota}
\nc{\qed}{\hskip 3em \hbox{\BOX} \vskip 2ex}

\nc{\real}{{\rm I \! R}}
\nc{\Z}{{\sf Z \!\!\! Z}}
\nc{\complex}{{\rm C\!\!\! {\sf I}\,\,}}
\def\bigid{\leavevmode\hbox{\small1\kern-3.8pt\normalsize1}}
\def\id{\leavevmode\hbox{\small1\kern-3.3pt\normalsize1}}
\nc{\slask}{\!\!\!/}
\nc{\bis}{{\prime\prime}}
\nc{\pa}{\partial}
\nc{\na}{\nabla}
\nc{\ra}{\rangle}
\nc{\la}{\langle}
\nc{\goto}{\rightarrow}
\nc{\swap}{\leftrightarrow}

\nc{\EE}[1]{ \mbox{$\cdot10^{#1}$} }
\nc{\abs}[1]{\left|#1\right|}
\nc{\at}[2]{\left.#1\right|_{#2}}
\nc{\norm}[1]{\|#1\|}
\nc{\abscut}[2]{\Abs{#1}_{\scriptscriptstyle#2}}
\nc{\vek}[1]{{\rm\bf #1}}
\nc{\integral}[2]{\int\limits_{#1}^{#2}}
\nc{\inv}[1]{\frac{1}{#1}}
\nc{\dd}[2]{{{\partial #1}\over{\partial #2}}}
\nc{\ddd}[2]{{{{\partial}^2 #1}\over{\partial {#2}^2}}}
\nc{\dddd}[3]{{{{\partial}^2 #1}\over
	{\partial #2 \partial #3}}}
\nc{\dder}[2]{{{d #1}\over{d #2}}}
\nc{\ddder}[2]{{{d^2 #1}\over{d {#2}^2}}}
\nc{\dddder}[3]{{d^2 #1}\over
	{d #2 d #3}}
\nc{\dx}[1]{d\,^{#1}x}
\nc{\dy}[1]{d\,^{#1}y}
\nc{\dz}[1]{d\,^{#1}z}
\nc{\dl}[1]{\frac{d\,^{#1}l}{(2\pi)^{#1}}}
\nc{\dk}[1]{\frac{d\,^{#1}k}{(2\pi)^{#1}}}
\nc{\dq}[1]{\frac{d\,^{#1}q}{(2\pi)^{#1}}}

\nc{\cc}{\mbox{$c.c.$ }}
\nc{\hc}{\mbox{$h.c.$ }}
\nc{\cf}{cf.\ }
\nc{\erfc}{{\rm erfc}}
\nc{\Tr}{{\rm Tr\,}}
\nc{\tr}{{\rm tr\,}}
\nc{\pol}{{\rm pol}}
\nc{\sign}{{\rm sign}}
\nc{\bfT}{{\bf T }}
\def\eV{{\rm\ eV}}
\def\GeV{{\rm\ GeV}}
\def\MeV{{\rm\ MeV}}

\def\TeV{{\rm\ TeV}}

\nc{\cA}{{\cal A}}
\nc{\cB}{{\cal B}}
\nc{\cD}{{\cal D}}
\nc{\cE}{{\cal E}}
\nc{\cG}{{\cal G}}
\nc{\cH}{{\cal H}}
\nc{\cL}{{\cal L}}
\nc{\cO}{{\cal O}}
\nc{\cT}{{\cal T}}
\nc{\cN}{{\cal N}}
\nc{\rvac}[1]{|{\cal O}#1\rangle}
\nc{\lvac}[1]{\langle{\cal O}#1|}
\nc{\rvacb}[1]{|{\cal O}_\beta #1\rangle}
\nc{\lvacb}[1]{\langle{\cal O}_\beta #1 |}
\nc{\bb}{\bar{\beta}}
\nc{\bt}{\tilde{\beta}}
\nc{\ctH}{\tilde{\cal H}}
\nc{\chH}{\hat{\cal H}}
\nc{\1}{\aa}
\nc{\2}{\"{a}}
\nc{\3}{\"{o}}
\nc{\4}{\AA}
\nc{\5}{\"{A}}
\nc{\6}{\"{O}}
\nc{\al}{\alpha}
\nc{\g}{\gamma}
\nc{\Del}{\Delta}
\nc{\eps}{\epsilon}
\nc{\lam}{\lambda}
\nc{\om}{\omega}
\nc{\Om}{\Omega}
\nc{\ve}{\varepsilon}
\nc{\mn}{{\mu\nu}}
\nc{\ka}{\kappa}
\nc{\vp}{\varphi}

\nc{\advp}[3]{{\it  Adv.\ in\ Phys.\ }{{\bf #1} {(#2)} {#3}}}
\nc{\annp}[3]{{\it  Ann.\ Phys.\ (N.Y.)\ }{{\bf #1} {(#2)} {#3}}}
\nc{\apl}[3]{{\it  Appl. Phys. Lett. }{{\bf #1} {(#2)} {#3}}}
\nc{\apj}[3]{{\it  Ap.\ J.\ }{{\bf #1} {(#2)} {#3}}}
\nc{\apjl}[3]{{\it  Ap.\ J.\ Lett.\ }{{\bf #1} {(#2)} {#3}}}
\nc{\app}[3]{{\it Astropart.\ Phys.\ }{{\bf #1} {(#2)} {#3}}}
\nc{\cmp}[3]{{\it  Comm.\ Math.\ Phys.\ }{{ \bf #1} {(#2)} {#3}}}
\nc{\cqg}[3]{{\it  Class.\ Quant.\ Grav.\ }{{\bf #1} {(#2)} {#3}}}
\nc{\epl}[3]{{\it  Europhys.\ Lett.\ }{{\bf #1} {(#2)} {#3}}}
\nc{\ijmp}[3]{{\it Int.\ J.\ Mod.\ Phys.\ }{{\bf #1} {(#2)} {#3}}}
\nc{\ijtp}[3]{{\it Int.\ J.\ Theor.\ Phys.\ }{{\bf #1} {(#2)} {#3}}}
\nc{\jmp}[3]{{\it  J.\ Math.\ Phys.\ }{{ \bf #1} {(#2)} {#3}}}
\nc{\jpa}[3]{{\it  J.\ Phys.\ A\ }{{\bf #1} {(#2)} {#3}}}
\nc{\jpc}[3]{{\it  J.\ Phys.\ C\ }{{\bf #1} {(#2)} {#3}}}
\nc{\jap}[3]{{\it J.\ Appl.\ Phys.\ }{{\bf #1} {(#2)} {#3}}}
\nc{\jpsj}[3]{{\it J.\ Phys.\ Soc.\ Japan\ }{{\bf #1} {(#2)} {#3}}}
\nc{\lmp}[3]{{\it Lett.\ Math.\ Phys.\ }{{\bf #1} {(#2)} {#3}}}
\nc{\mpl}[3]{{\it  Mod.\ Phys.\ Lett.\ }{{\bf #1} {(#2)} {#3}}}
\nc{\ncim}[3]{{\it  Nuov.\ Cim.\ }{{\bf #1} {(#2)} {#3}}}
\nc{\np}[3]{{\it  Nucl.\ Phys.\ }{{\bf #1} {(#2)} {#3}}}
\nc{\npps}[3]{{\it  Nucl.\ Phys.\ Proc.\ Suppl.\ }{{\bf #1} {(#2)} {#3}}}
\nc{\pr}[3]{{\it Phys.\ Rev.\ }{{\bf #1} {(#2)} {#3}}}
\nc{\pra}[3]{{\it  Phys.\ Rev.\ A\ }{{\bf #1} {(#2)} {#3}}}
\nc{\prb}[3]{{\it  Phys.\ Rev.\ B\ }{{{\bf #1} {(#2)} {#3}}}}
\nc{\prc}[3]{{\it  Phys.\ Rev.\ C\ }{{\bf #1} {(#2)} {#3}}}
\nc{\prd}[3]{{\it  Phys.\ Rev.\ D\ }{{\bf #1} {(#2)} {#3}}}
\nc{\prl}[3]{{\it Phys.\ Rev.\ Lett.\ }{{\bf #1} {(#2)} {#3}}}
\nc{\pl}[3]{{\it  Phys.\ Lett.\ }{{\bf #1} {(#2)} {#3}}}
\nc{\prep}[3]{{\it Phys.\ Rept.\ }{{\bf #1} {(#2)} {#3}}}
\nc{\prsl}[3]{{\it Proc.\ R.\ Soc.\ London\ }{{\bf #1} {(#2)} {#3}}}
\nc{\ptp}[3]{{\it  Prog.\ Theor.\ Phys.\ }{{\bf #1} {(#2)} {#3}}}
\nc{\ptps}[3]{{\it  Prog\ Theor.\ Phys.\ suppl.\ }{{\bf #1} {(#2)} {#3}}}
\nc{\physa}[3]{{\it  Physica\ A\ }{{\bf #1} {(#2)} {#3}}}
\nc{\physb}[3]{{\it  Physica\ B\ }{{\bf #1} {(#2)} {#3}}}
\nc{\phys}[3]{{\it Physica\ }{{\bf #1} {(#2)} {#3}}}
\nc{\rmp}[3]{{\it  Rev.\ Mod.\ Phys.\ }{{\bf #1} {(#2)} {#3}}}
\nc{\rpp}[3]{{\it Rep.\ Prog.\ Phys.\ }{{\bf #1} {(#2)} {#3}}}
\nc{\sjnp}[3]{{\it Sov.\ J.\ Nucl.\ Phys.\ }{{\bf #1} {(#2)} {#3}}}
\nc{\spjetp}[3]{{\it Sov.\ Phys.\ JETP\ }{{\bf #1} {(#2)} {#3}}}
\nc{\yf}[3]{{\it Yad.\ Fiz.\ }{{\bf #1} {(#2)} {#3}}}
\nc{\zetp}[3]{{\it Zh.\ Eksp.\ Teor.\ Fiz.\  }{{\bf #1}  {(#2)} {#3}}}
\nc{\zp}[3]{{\it Z.\ Phys.\ }{{\bf #1} {(#2)} {#3}}}
\nc{\ibid}[3]{{\sl ibid.\ }{{\bf #1} {#2} {#3}}}
\nc{\rf}[1]{(\ref{#1})}
\nc{\nn}{\nonumber \\*}
\nc{\bfB}{\bf{B}}
\nc{\bfv}{\bf{v}}
\nc{\bfx}{\bf{x}}
\nc{\bfy}{\bf{y}}
\nc{\vx}{\vec{x}}
\nc{\vy}{\vec{y}}
\nc{\oB}{\overline{B}}
\nc{\oI}{\overline{I}}
\nc{\oR}{\overline{R}}
\nc{\rar}{\rightarrow}
\nc{\ti}{\times}
\nc{\slsh}{\hskip-5pt/}
\nc{\sm}{Standard~Model~}
\nc{\MP}{M_{\rm Pl}}
\nc{\tp}{t_{\rm Pl}}
\nc{\ave}{\bar{E}}

\nc{\eff}{{\rm eff}}
\nc{\kk}{\vek{k}}
\nc{\pp}{{\rm p}}
\nc{\ga}{g_{a\gamma}}
\nc{\vv}{\\}
\nc{\eee}{{\bf E}}
\nc{\bbb}{{\bf B}}
\nc{\qcd}{T_{\rm QCD}}
\nc{\G}{\rm \ G}
\def\vec#1{{\bf #1}}
\def\vv{\vskip-2pt}

\def\ell{e^{c}LL}

\begin{document}
{\title{\vskip-2truecm{\hfill {{\small HIP-2002-15/TH\\
	\hfill \\
	}}\vskip 1truecm}
{\LARGE Analytical and numerical properties of Affleck-Dine condensate formation}}
\vspace{-.2cm}
{\author{
{\sc Asko Jokinen$^{1}$}
\\
{\sl\small Helsinki Institute of Physics, P.O. Box 64, FIN-00014 University of Helsinki, FINLAND}}}
\maketitle
\vspace{1cm}
\begin{abstract}
\noindent
We present analytical and numerical properties of the coherently rotating Affleck-Dine condensate formed along MSSM flat directions with the radiative corrections to the mass terms included. We analyse the pressure $p$ of the condensate, which is known to be negative if the potential grows slower than the field squared. We show that ellipticity of the orbit of the rotating field also affects the pressure. For circularly orbiting field the $p=0$ while the smallest negative value, corresponding to the most unstable configuration, is obtained for a coherently oscillating field. The AD condensate is known to fragment into non-topological solitons called Q-balls. We also study equilibration of Q-ball distribution resulting from the fragmentation of the condensate analytically. We find that equilibration is dependent on the energy-to-charge ratio $x$ of the condensate. We find the allowed range of $x$ numerically and deduce that equilibration is likely to happen.
\end{abstract}
\vfil
\footnoterule
{\small $^1$asko.jokinen@helsinki.fi}

\thispagestyle{empty}
\newpage
\setcounter{page}{1}

\section{Introduction}
The Affleck-Dine (AD) mechanism \cite{ad} provides a model for generating the observed baryon asymmetry of the universe in the framework of supersymmetry. In this scenario some squarks and/or sleptons acquire a large vacuum expectation value (VEV) along a flat direction (AD field) of the scalar potential of the MSSM during inflationary epoch. A baryon/lepton number violating operator is induced by new physics at a high scale together with a large C and CP violating phase, which is provided by the initial VEV along the flat direction. Together with the out of equilibrium conditions after inflation the three requirements for the generation of baryon asymmetry are satisfied \cite{sakh}. The AD field starts oscillating\footnote{Strictly speaking AD field is rotating if the condensate has a non-zero charge. Pure oscillation has charge zero. But in the literature the term oscillation has become standard.} once its effective mass exceeds the Hubble expansion rate $H$. At the same time a baryon/lepton number violating operator produces a torque towards the phase direction which leads to a spiral motion of the VEV in the complex field space. This results in a baryon/lepton asymmetry in a form of spatially homogeneous condensate once the comoving number density of the AD particles is frozen at sufficiently late times.

The dynamics after the AD condensate formation is determined by the mass of the flat direction. This is essentially given by the soft SUSY breaking mass terms of the MSSM since the Hubble induced mass term is negligible at this stage. When radiative corrections are taken into account the mass term typically tends to grow slower than quadratically \cite{ksad,bbb1}. This induces in the condensate a negative pressure and makes it unstable with respect to spatial perturbations. Due to the instabilities the AD condensate fragments into lumps of charge, which relax into Q-balls. Q-ball is a field configuration which minimizes the energy with non-zero charge \cite{cole}. The fragmentation process is highly non-linear and has to be solved numerically. AD condensate fragmentation has been extensively studied in \cite{kskm1,kskm2,kskm3,ejmv,muvi3} where it was noted that the original charge of the AD condensate is stored in the Q-balls. It was also noted that the relative amount of Q-balls and anti-Q-balls is determined by the dimensionless ratio of energy per charge times the scalar mass, denoted as $x$ in this paper. For $x>1$ there appear anti-Q-balls in addition to Q-balls. If $x\gg 1$ the distribution of Q-balls approaches a thermal one \cite{ejmv,muvi3}. Therefore it is important to know the energy of the condensate in order to study the fragmentation process.

In this paper we calculate numerically the properties of the AD condensate: charge density in the comoving volume, energy-to-charge ratio $x$ and the pressure-to-energy ratio $w$ which gives the equation of state. When the charge density is known we can estimate the resulting charge-to-entropy ratio, which is known from nucleosynthesis calculations. This has been studied in \cite{drt,anis}. However, $x$ and $w$ have not been calculated previously. We find that generically $|x|>1.1$ in gravity mediated SUSY breaking models. For gauge mediated models the situation is more involved because the mass is almost a constant at the high energy scale. This effectively leads to a larger negative pressure which makes the energy of the condensate grow rapidly. This effect is also seen in the gravity mediated case but is much milder.

These same considerations apply to any extension of the MSSM. For the extensions there can exist flat directions which carry charges different from baryon or lepton number but the same considerations apply provided that their radiative corrections give a mass term that again grows slower than quadratically. Q-balls, which do not carry baryon or lepton number, could act as self-interacting dark matter \cite{spst,kustein,ejmv2}.

This paper is organized as follows. In Sec. II we review the main features of AD mechanism \cite{drt}, calculate the pressure of a rotating condensate by extending the analysis of \cite{turner} for oscillating condensates, review the relevant properties of Q-balls and re-analyze the reaction rates to show that for large $x$ a thermal equilibrium is a generic feature. In Sec. III we give the details and results of the numerical analysis. In Sec. IV we give the conclusions.

\section{General properties of AD mechanism}
\subsection{The evolution of flat directions}
The potential for the flat direction is induced by the soft SUSY breaking mass terms, the A-terms of the MSSM and by the finite energy SUSY breaking due to the supergravity coupling of the flat direction to the inflaton superfield in the early universe. All these effects result in a potential of the form \cite{drt}
\be{pot1}
V(\Phi) = V_m(\Phi) - c_HH^2|\Phi|^2 + \left(\frac{Am_{3/2}+aH}{dM_p^{d-3}}\,\lam\Phi^d + h.c.\right) + \frac{|\lam|^2}{M_p^{2(d-3)}}\,|\Phi|^{2(d-1)},
\ee
where $c_H\sim 1$, $M_p\sim 10^{18}\GeV$ is a high mass scale which here has been chosen as the reduced Planck mass, $d$ is the dimension of the non-renormalizable operator that lifts the flat direction (we consider the cases $d=4,\,5,\,6,\,7$) and $\lam$, $A$ and $a$ are in general complex constants $\cO(1)$. Here we make the difference between D- and F-term inflation \cite{lyri}: in the D-term case the Hubble induced A-term is absent, \ie $a=0$, and in the F-term case $a\sim 1$. $V_m(\Phi)$ is the mass term of the flat direction which, including the radiative corrections, in case of gravity mediated SUSY breaking \cite{bbb1} is
\be{mgrav}
V_m(\Phi) = m_{3/2}^2\left[1+K\log\left(\frac{|\Phi|^2}{M^2}\right)\right]|\Phi|^2,
\ee
where $m_{3/2}\sim 1\TeV$ is the gravitino mass, $K\sim -0.1\ldots -0.01$ when considering unstable flat directions and $M\sim (M_p^{d-3} m_{3/2}/|\lam|)^{1/(d-2)}$ is the associated renormalization scale \cite{ejm}. In the case of gauge mediated SUSY breaking \cite{ksad}
\be{mgauge}
V_m(\Phi) = m_{\Phi}^4\log\left(1+\frac{|\Phi|^2}{m_{\Phi}^2}\right),
\ee
where $m_{\Phi}\sim 1\ldots 100\TeV$; in this case the gravitino mass $m_{3/2}<1\GeV$. One should note that in order to have an AD condensate formation the parameters of the A-term are constrained: in the gravity mediated case $|A|<d$ and in the gauge mediated case $|A|m_{3/2}\lsim 10^{-4},\,10^{-7}m_{\Phi}$ when $d=4,\,6$. If the condition in the gauge mediated case is not fulfilled, one just has to take the gravity mediated effects, \eq{mgrav}, into account and the behaviour becomes of the gravity mediated type at the time of condensate formation.

We could also include thermal corrections to the potential, \eq{pot1}, as has been done recently \cite{kskm2,anis,ace,andi}. Their effect is of the same form as \eqs{mgrav}{mgauge} if one replaces either $m_{3/2}$ or $m_{\Phi}$ by the temperature $T\gg m_{3/2},\,m_{\Phi}$. In principle one would have to include also the gravitational corrections in gauge mediated case, too, but since all the relevant quantities of the AD condensate are determined by the most dominant mass term we only have to pay attention to one mass term at a time in the simulations.

The main contribution to the evolution of the flat direction comes from the homogeneous mode whose equation of motion is
\be{eqm1}
\ddot\Phi + 3H\dot\Phi + \dd{V}{\Phi^*} = 0.
\ee
It is easier to analyze the behaviour of the equations of motion by parameterizing the field by $\Phi=\frac{1}{\sqrt{2}}\,\phi\,e^{i\theta}$, for numerical purposes in Section 3 the parameterization $\Phi=\frac{1}{\sqrt{2}}(\phi_1+i\phi_2)$ is used. Then one obtains
\bea{eqm2}
\ddot\phi + 3H\dot\phi - \dot\theta^2\phi + \dd{V}{\phi} = 0, \nn \ddot\theta + \left(3H+\frac{2\dot\phi}{\phi}\right)\,\dot\theta + \frac{1}{\phi^2}\,\dd{V}{\theta} = 0.
\eea

In order to understand qualitatively the behaviour of the equations of motion \eq{eqm2} we find the maxima and minima of the potential, which are achieved with the phase values
\bea{theta1}
\theta = -\frac{1}{d}\arctan\left(\frac{|A|m_{3/2}\sin(\theta_A+\theta_{\lam})+|a|H\sin(\theta_a+\theta_{\lam})}{|A|m_{3/2}\cos(\theta_A+\theta_{\lam})+|a|H\cos(\theta_a+\theta_{\lam})}\right)+\frac{2\,n+b}{d}\,\pi,
\eea
where $b=0$ for a maximum and $b=1$ for a minimum.\footnote{Note that this is valid only if the denumerator is positive. If the denumerator is negative one has to switch $b=0$ for a minimum and $b=1$ for a maximum.} This can be simplified for $H\gg m_{3/2}$ for which the phase minimum is at $d\theta=-(\theta_a+\theta_{\lam})+(2n+1)\pi$, while for $H\ll m_{3/2}$ the minimum is at $d\theta=-(\theta_A+\theta_{\lam})+(2n+1)\pi$. For $H\gg m_{3/2},\,m_{\Phi}$ one can solve the minimum for $\phi$ approximately as
\be{phi1}
\phi_{\min} = \sqrt{2}\left[\frac{M_p^{d-3}\,H}{2\,|\lam|\,(d-1)}\,\left(-f(\theta) + \sqrt{f(\theta)^2+4(d-1)c_H}\right)\right]^{1/(d-2)},
\ee
where
\be{ftheta}
f(\theta) = |A|\frac{m_{3/2}}{H}\cos(d\theta+\theta_A+\theta_{\lam}) + |a|\cos(d\theta+\theta_a+\theta_{\lam}).
\ee
These will be needed for numerical simulations as the initial conditions.

Now we can analyze the AD condensate formation qualitatively by noting that the AD potential is analogous to potentials giving rise to a first order phase transition in the $\phi$ direction and in the F-term case there is also a second order transition in the $\theta$ direction. However, in this case there appears no tunneling. In this case the Hubble parameter determines the behaviour of the potential, not the temperature as usual. During inflation the AD field $\Phi$ settles to one of the symmetry breaking minima in \eqs{theta1}{phi1} in the F-term case. In the D-term case the phase $\theta$ stays random and only $\phi$ acquires a minimum value \eq{phi1} \cite{drt}. After inflation the Universe becomes matter dominated by virtue of inflaton oscillations and the Hubble parameter reads $H=2/(3t)$ so that the minimum, \eq{phi1}, evolves towards a lower energy scale. $\phi$ evolves close to a fixed point value of the equations of motion \eq{eqm2} \cite{drt}, which is slightly larger than the minimum (see \eq{phi1}), and given by
\be{phifix}
\phi=\left(1+\frac{9(d-3)}{4c_H(d-2)^2}\right)^{\frac{1}{2(d-2)}}\phi_{\min}(t).
\ee

Eventually the Hubble mass term becomes equal to the mass term coming from soft SUSY breaking, \eqs{mgrav}{mgauge}, $H_{pt}^2=V_m(\Phi)/|\Phi|^2$. At this scale the phase transition from $\phi=\phi_{\min}$, \eq{phi1}, to $\phi=0$ occurs and the AD field $\Phi$ starts to rotate in the pit of the symmetry breaking minimum. For the gravity mediated case the phase transition happens when $H_{pt}\sim m_{3/2}$. In the gauge mediated case it happens at $H_{pt}\sim 2\ldots 5\,m_{\Phi}|\lam|^{1/(d-1)}(m_{\Phi}/M_p)^{(d-3)/(d-1)}$, which can be calculated by noting that the logarithm is on the range $10\ldots 100$ for a large range of $\phi$ and $m_{\Phi}$. After a while the symmetry breaking minimum vanishes while the AD field continues its rotation in a spiraling orbit around $\Phi=0$. This is what charges up the condensate, the charge density of which is defined by
\be{charge1}
q=\frac{1}{i}(\dot\Phi\Phi^*-\Phi\dot\Phi^*)=\dot\theta\phi^2.
\ee
The charge density in the co-moving volume asymptotes to a constant when $H\sim 0.1H_{pt}$. \fig{plotreim} shows the condensate formation. More on this in Section 3.

The subsequent behaviour is determined by the mass term of the field $\Phi$. In the original AD scenario \cite{ad,drt} the mass term was given by $V_m(\Phi)=m^2|\Phi|^2$ and the condensate decayed into $\Phi$-quanta. The evolution is however totally different if the potential grows slower than $\Phi^2$ as is the case considered in this paper. Then the condensate fragments into non-topoligical colitons called Q-balls \cite{ksad,bbb1}.

\subsection{Analytical properties of the AD condensate}
The charge density, \eq{charge1},is calculated using the second equation of \eq{eqm2} in the form $d/dt(\dot\theta\phi^2R^3)=-R^3\,\partial V/\partial\theta$, where $R=R_0(t/t_0)^{2/3}$ is the scale factor of the Universe, giving approximately
\be{charge}
q\sim H^{-1}\dd{V_A}{\theta}\sim m_{3/2}\left(\frac{M_p^{d-3}H}{|\lam|}\right)^{\frac{2}{d-2}},
\ee
where $H\sim H_{pt}$ at the time of condensate formation.

When the condensate has been charged, we can approximately solve the equations of motion by neglecting all the non-renormalizable terms and the Hubble mass terms in \eq{pot1} and retain only the mass term in the equations of motion \eq{eqm2}:
\bea{eqm3}
\ddot\phi+3H\dot\phi-\dot\theta^2\phi+C\phi^{n-1}=0 \nn
\ddot\theta+\left(3H+\frac{2\dot\phi}{\phi}\right)\dot\theta=0.
\eea
The second equation of \eq{eqm3} merely gives conservation of charge density in the co-moving volume $Q=\dot\theta\phi^2R^3=const$. In the first equation of \eq{eqm3} the derivative of the mass term has been approximated by $C\phi^{n-1}$ (for the potential $V(\phi)\sim\phi^n$). For the gravity mediated case $C=2^{|K|}M^{2|K|}m_{3/2}^2$ and $n=2-2|K|$; for the gauge mediated case $C=2m_{\Phi}^4$ and $n=0$ can be used since the derivative of the potential has this behaviour for large $\phi$. There then exists a fixed point solution (neglecting terms of order $H^2$ which is consistent since these terms there neglected in \eq{eqm3}, too) with
\be{fix}
\phi=\left(\frac{Q^2}{C}\right)^{\frac{1}{n+2}}\,R^{-\frac{6}{n+2}}.
\ee
This behaviour is exactly the same as was obtained for coherently oscillating condensates in \cite{turner}. \eq{fix} gives $\phi\sim R^{-3/(2-|K|)}$ for the gravity mediated case and $\phi\sim R^{-3}$ for the gauge mediated case. The usual method of analyzing the stability of the fixed point by calculating the eigenvalues of the matrix of the linearized equations does not work, because \eq{eqm3} are non-autonomous differential equations.

Another feature of coherent condensates with potential growing slower than $\Phi^2$ is that the condensate has a negative pressure \cite{bbb1,turner,mcdon}. As a consequence it becomes unstable since the energy in a co-moving volume increases, as can be seen from the continuity equation
\be{cont}
\dot\rho + 3H(\rho+p) - \dd{V}{t}= 0,
\ee
where $p=|\dot\Phi|^2-V(\Phi)$ and $\rho=|\dot\Phi|^2+V(\Phi)$ and the partial derivative of potential with respect to time appears, because the potential depends explicitly on time through $H$. However, after condensate formation that term is negligible, since $H^2\ll V(\Phi)/|\Phi|^2$. The equation of state for coherent condensates can be found by averaging $w=p/\rho=2|\dot\Phi|^2/\rho-1$ over a cycle of rotation as done for oscillating condensates in \cite{turner}, so we want to calculate the time average of $\frac{1}{T}\int_0^T dt\,|\dot\Phi|^2/\rho$ \cite{turner}. The calculation can be done by noting that $|\dot\Phi|dt=d|\Phi|$ is the arc length in the $\Phi$ space. Now we get for $w$
\be{w1}
w = 2\,\frac{\integral{C}{} d|\Phi|\,\left(1-\frac{V}{\rho}\right)^{1/2}}{\integral{C}{} d|\Phi|\,\left(1-\frac{V}{\rho}\right)^{-1/2}} - 1,
\ee
where we are now integrating over the orbit of the AD-field $\Phi$. This is exactly the same as was the case for pure oscillation in \cite{turner} except that now the integration is over a different orbit and the energy density is equal to $\rho=V_{\max}+V_{\min}$ (for pure oscillation $V_{\min}=0$). Usually the orbit is quite close to an elliptical orbit with a semimajor axis $A$ and a semiminor axis $B$ giving the arc length as
\be{arc}
d|\Phi| = \sqrt{1+\frac{B^2\varphi_1^2}{A^4\left(1-\frac{\varphi_1^2}{A^2}\right)}}\,\frac{d\varphi_1}{\sqrt{2}}
\ee
and the integration limits are $0\leq\varphi_1\leq A$ ($B\leq A$). \eq{w1} cannot be given in a closed form except in a few cases: for circular orbit $B=A$ from which follows that $V_{\max}=V_{\min}$ and $\rho=2V$, giving the result $w=0$; for pure oscillation $B=0$ (no charge in the condensate in this case) when $V(\Phi)\sim |\Phi|^n$, giving $w=(n-2)/(n+2)$ \cite{turner}; and if $n=2$ for all values of $0\leq B\leq A$, we get $w=0$.

Let us generalize the result of \cite{turner} by considering elliptical orbits in a potential $V(\Phi)\sim |\Phi|^n$. We can simplify \eq{w1} to depend only on the ratio of semi-major axes $\eps=B/A$ (actually $\eps=\sqrt{1-e^2}$ where $e$ is the eccentricity of the ellipse)
\be{w2}
w = 2\,\frac{\integral{0}{1} dx\left(1+\frac{\eps^2x^2}{1-x^2}\right)^{1/2}\,\left[1-\frac{\left(x^2+\eps^2(1-x^2)\right)^{n/2}}{1+\eps^n}\right]^{1/2}}{\integral{0}{1} dx\left(1+\frac{\eps^2x^2}{1-x^2}\right)^{1/2}\,\left[1-\frac{\left(x^2+\eps^2(1-x^2)\right)^{n/2}}{1+\eps^n}\right]^{-1/2}} - 1,
\ee
where a change of variable $\varphi_1=Ax$ was done. When $\eps$ goes from 0 to 1, the orbit changes from pure oscillation to circular rotation. At the same $w$ changes from $(n-2)/(n+2)$ to zero. Thus for $n<2$ the pressure is negative. We have plotted in \fig{plotpreeneps}(a) the result of the numerical integration of \eq{w2} in the gravity mediated case. For pure oscillation the pressure has been calculated to be $p\approx\frac{K}{2}\rho$ \cite{mcdon}, which is negative when $K<0$.

\bfig
\leavevmode
\centering
\vspace*{4cm}
\includegraphics{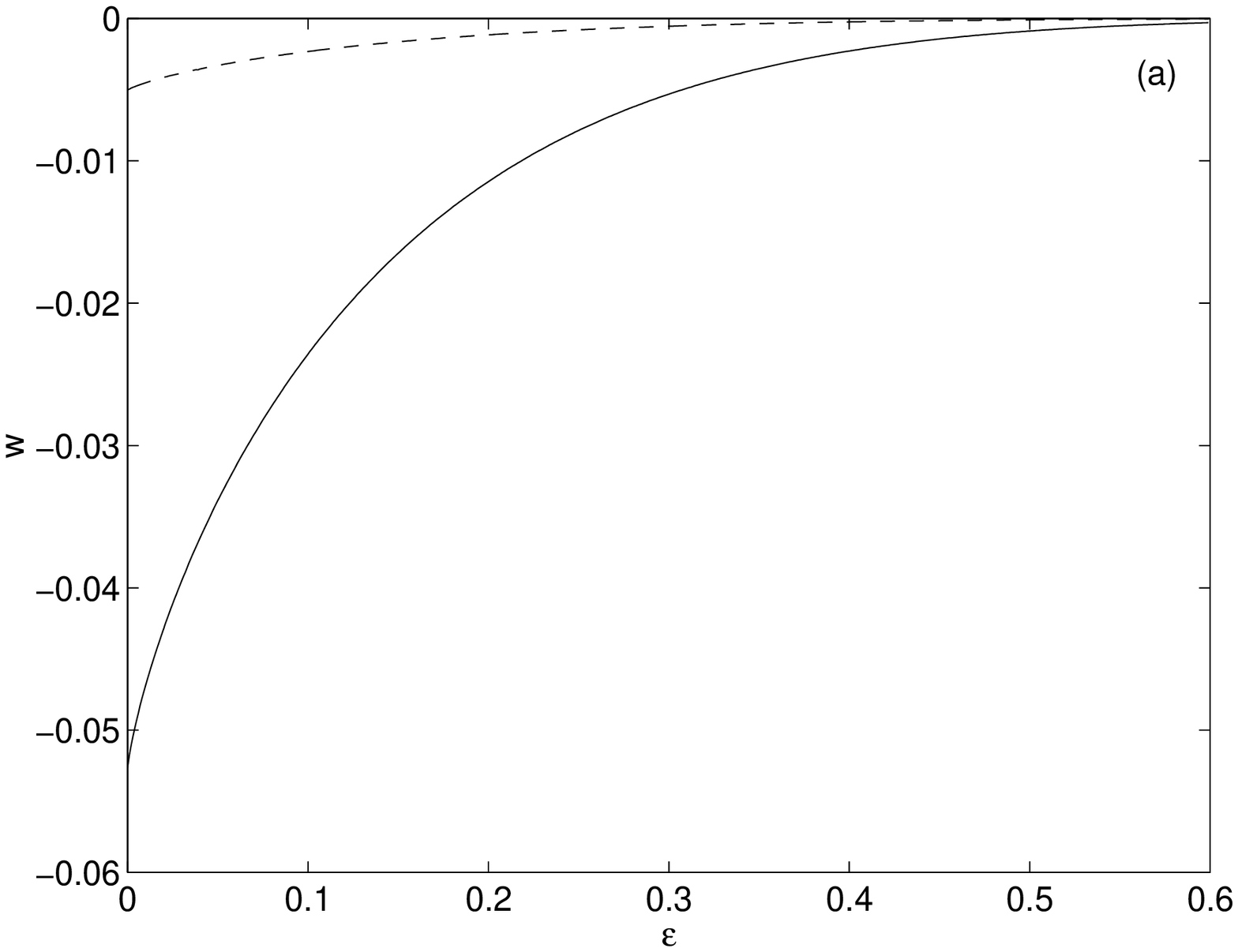}
\includegraphics{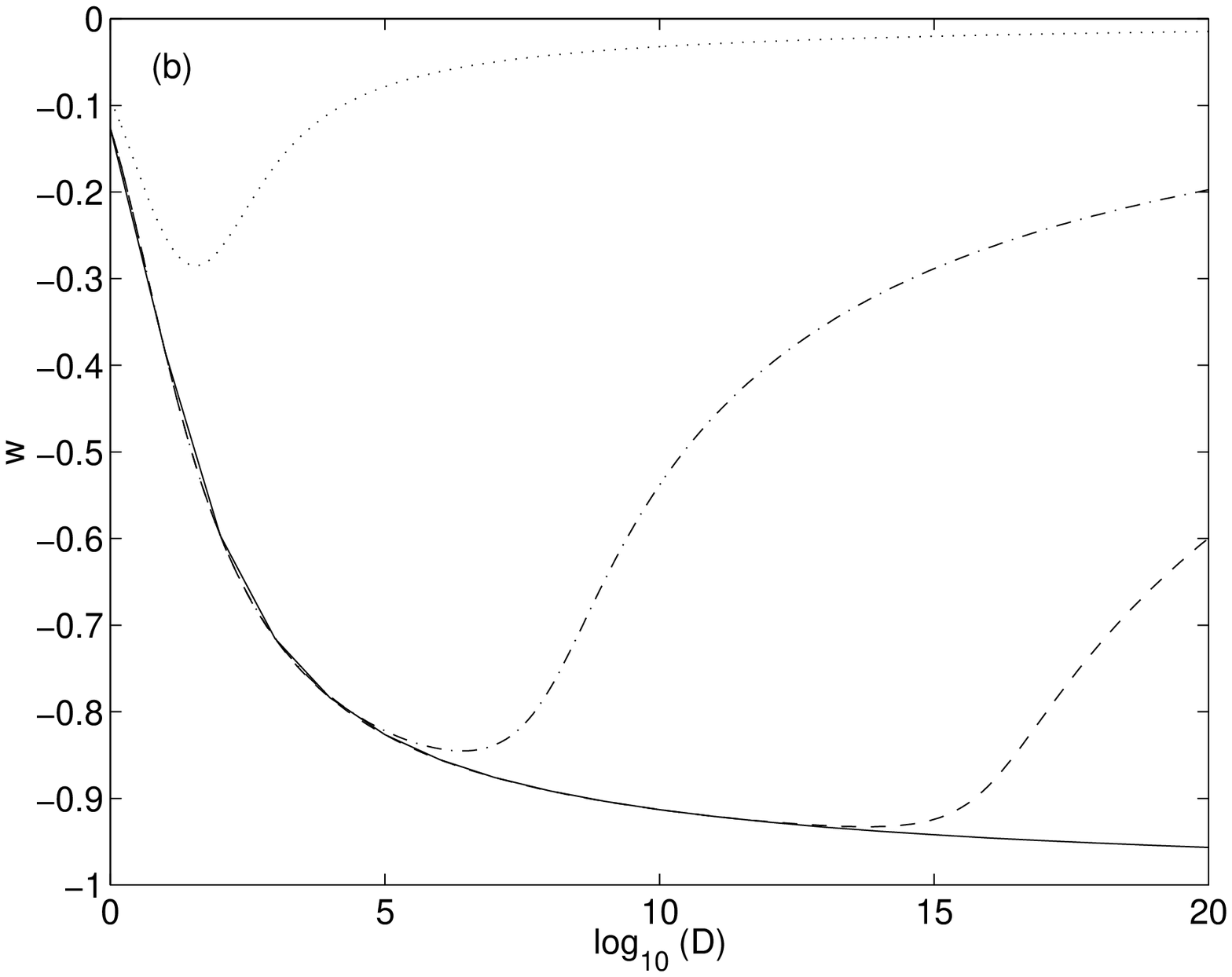}
\caption{\small Pressure-to-energy ratio, $w$, plots (a) gravity mediated case for $K=-0.1$ (solid) and $K=-0.01$ (dashed), (b) gauge mediated case plotted against $D=\phi_{\max}^2/(2m_{\Phi}^2)$ for $\eps=0,\,10^{-8},\,10^{-4},\,0.1$ with solid, dashed, dash-dotted and dotted lines.}\label{plotpreeneps}
\efig

For the logarithmic mass term of the gauge mediated scenario, \eq{mgauge}, the calculation of pressure is more involved. If we approximate the potential as a constant, as was done in \cite{ksad}, we get for pure oscillation $w=-1$ by putting $n=0$, thus yielding the smallest value of $w$ possible. With the logarithmic mass term we can evaluate $w$ numerically by making the approximation that the orbit is elliptical. (We shall see in Section 3 that the orbit of the AD field $\Phi$ is not really elliptical but looks more like a rotating trefoil.) Thus we obtain
\be{w3}
w = 2\frac{\integral{0}{1} dx\left(1+\frac{\eps^2x^2}{1-x^2}\right)^{1/2}\,\left[1-\frac{\log\left(1+D(x^2+\eps^2(1-x^2))\right)}{\log(1+D)+\log(1+D\eps^2)}\right]^{1/2}}{\integral{0}{1} dx\left(1+\frac{\eps^2x^2}{1-x^2}\right)^{1/2}\,\left[1-\frac{\log\left(1+D(x^2+\eps^2(1-x^2))\right)}{\log(1+D)+\log(1+D\eps^2)}\right]^{-1/2}} -1,
\ee
where $D=A^2/(2m_{\Phi}^2)$. In \fig{plotpreeneps} we plot the numerical results of \eq{w3} for several values $D$.

We have seen that for potentials that grow slower than $|\Phi^2|$, the pressure is negative unless the orbit of the AD field is circular. In an expanding Universe a circular orbit for a spatially homogeneous condensate is impossible because of the dissipation. Therefore the energy density of the AD condensate in the co-moving volume increases. Thus the condensate is unstable and the most unstable orbit is purely oscillating orbit, where the charge of the condensate is zero. This feature was also seen in the simulation of Kasuya and Kawasaki \cite{kskm3}, where they showed that there exist more unstable modes for pure oscillation compared to the circular orbit. This is the instability that causes the condensate to fragment. While it would be natural to expect that the condensate decays into $\Phi$-quanta, for potentials growing slower than $\Phi^2$ the minimum energy configuration is an ensemble of Q-balls.

\subsection{Q-balls}
Q-balls have the generic form
\be{qball}
\Phi({\bf{x}},t) = \frac{1}{\sqrt{2}}\phi(r)e^{i\omega t},
\ee
where $\phi(r)$ is a decreasing function of $r$. The value of rotation velocity $\omega$ and the form of $\phi(r)$ depend on details of the potential. Note that a Q-ball is rotating circularly in the field space. Unlike for the coherent condensate, this is now possible because the Q-ball is spatially inhomogeneous. The different kinds of Q-balls considered in the literature can either have a narrow well-defined edge, in which case they are called thin-wall Q-balls \cite{cole}, or their boundaries are not localized, in which case they are called thick-wall Q-balls \cite{k}. In this paper we concentrate only on thick-wall Q-balls.

Thick-wall Q-balls arise in supersymmetric theories with gauge and gravity mediated supersymmetry breaking \cite{dks,bbb2}. In the gauge mediated case the field value inside the Q-ball, energy and radius are are given by \cite{dks}
\bea{qgauge}
\phi_0 &=& \frac{m_{\Phi}}{\sqrt{\pi}}Q^{\frac{1}{4}} \nn
E &=& \frac{4\pi\sqrt{2}}{3}m_{\Phi}Q^{\frac{3}{4}} \nn
R_Q &=& \frac{1}{\sqrt{2}m_{\Phi}}Q^{\frac{1}{4}},
\eea
where the profile of the field $\phi$ is well approximated by a kink-solution \cite{muvi2}. In the gravity mediated case the corresponding quantities are \cite{bbb2}
\bea{qgrav}
\phi(r) &=& \phi_0e^{-\frac{r^2}{R^2}} \nn
E &=& m_{3/2}Q \nn
R_Q &=& \frac{1}{\sqrt{|K|}m_{3/2}}.
\eea
There is also the possibility having mixed Q-balls with both mass terms \eqs{mgrav}{mgauge} of gravity and gauge mediated cases together \cite{kskm4}.

The energy (mass) of the Q-ball has to be less than the mass of $Q$ free scalars \ie $E<m|Q|$ in order to be stable with respect to decay into free scalars. For this reason the AD condensate, which is a spatially homogeneous rotating field, fragments into Q-balls and not to free scalars. The interesting cases for us are the thick-walled Q-balls.

\subsection{Condition for thermal equilibrium}
In \cite{ejmv} it was noticed that for large values of energy-to-charge ratio $x$ (or, to be more precise, for large $|x|$) the end product of condensate fragmentation is an ensemble of both Q-balls and anti-Q-balls. The total charge they carry is much larger than the total charge of the condensate: $Q_++Q_-\gg Q_{tot}$, where $Q_+$ ($Q_-$) is the total charge carried by Q-balls (anti-Q-balls). This can be understood easily if both the total energy and charge are stored in Q-balls, as has been argued in \cite{kskm1,kskm2,kskm3,ejmv,muvi3}:
\bea{enchatot}
E_{tot} &=& \int d{\bf{p}}dQ (N_+({\bf{p}},Q)+N_-({\bf{p}},Q))\sqrt{p^2+m_Q^2} = E_++E_- \nn Q_{tot} &=& \int d{\bf{p}}dQ (N_+({\bf{p}},Q)+N_-({\bf{p}},Q))Q = Q_+-Q_-,
\eea
where $m_Q$ is given by \eqs{qgauge}{qgrav} and $N_{\pm}({\bf{p}},Q)$ are the number distribution functions of Q-balls, which at this point are arbitrary. We approximate the integrals in \eq{enchatot} by replacing $Q$ with an average charge $\bar{Q}>0$ of a Q-ball and $\sqrt{p^2+m_Q^2}$ with an average energy $\bar{E}$ to obtain $E_{\pm}\approx \bar{E}(N_++N_-)$ and $Q_{\pm}\approx \bar{Q}(N_+-N_-)$, where $N_{\pm}=\int d{\bf{p}}dQ N_{\pm}({\bf{p}},Q)$. Then we can approximate the energy-to-charge ratio $x$ by
\be{xappr}
x\approx \frac{\bar{E}}{m_{\bar{Q}}}\,\frac{m_{\bar{Q}}}{m\bar{Q}}\,\frac{Q_++Q_-}{Q_+-Q_-}.
\ee
The factor $\bar{E}/m_{\bar{Q}}$ can be interpreted as $\gamma=1/\sqrt{1-v^2}$. For the gravity mediated case, \eq{qgrav}, we see that if $x\gg 1$ then either $\gamma\gg 1$, which would make the Q-balls ultra-relativistic, or $Q_++Q_-\gg Q_+-Q_-=Q_{tot}$, which indicates that the total charges of Q-balls and anti-Q-balls are much larger than the original net condensate charge. Since it is unlikely that massive particles with $m\sim 100\GeV\ldots 100\TeV$ and $\bar{Q}\gg 1$ would be ultra-relativistic, there has to be a large charge asymmetry in Q-balls for $x\gg 1$. For the gauge mediated case \eq{qgauge} a large number of Q-balls and anti-Q-balls can be produced even if $x<1$, because \eq{xappr} contains a numerical factor $m_{\bar{Q}}/m\bar{Q}\ll 1$ for $\bar{Q}\gg 1$.

Since the number of Q-balls and anti-Q-balls is large, it is likely that there are numerous Q-ball collisions, which rapidly thermalize the distibution. The condition for this to happen is that the collision rate of Q-balls is larger than the Hubble rate,
\be{rate1}
\Gamma=n_{tot}\sigma v>H=\frac{2}{3t},
\ee
where $n_{tot}$ is the number density of Q-balls, $\sigma\approx \pi R_Q^2$ is the cross-section of a Q-ball collision and $v$ is the average velocity of a Q-ball. Using \eq{xappr} we can write the rate as
\be{rate2}
\Gamma\approx x\,\frac{m}{m_{\bar{Q}}}\,q_{tot}\pi R_{\bar{Q}}^2\,\frac{v}{\gamma},
\ee
where $q_{tot}=q_0(t_0/t)^2$ is the total charge density with $q_0$ the initial charge density of the condensate at time $t_0$. In order to get the most conservative bound for the collision rate we approximate the average charge of the Q-ball, $\bar{Q}$, with the maximum charge of a Q-ball, $Q_{\max}$. Fragmentation of a $d=4$ condensate was considered by Kasuya and Kawasaki both in the gravity mediated \cite{kskm2,kskm3} and gauge mediated case \cite{kskm1,kskm3}. They performed numerical simulations which begun at $t_0=2/(3m_{3/2})$ (grav.) and at $t_0=\sqrt{2}\phi_0/(3m_{\Phi}^2)$ (gauge) with initial charge density $q_0=m_{3/2}\phi_0^2$ and found that the maximum charge of the Q-balls formed is $Q_{\max}\approx 6\cdot 10^{-3}\phi_0^2m_{3/2}^{-2}$ (grav.) and $Q_{\max}\approx 6\cdot 10^{-4}\phi_0^4m_{\Phi}^{-4}$ (gauge). The Q-balls are formed at $t_f\sim 5\cdot 10^3m_{3/2}^{-1}$ (grav.) \cite{kskm2} and $t_f\sim 5\cdot 10^5m_{\Phi}^{-1}$ (gauge) \cite{kskm1}. Now \eq{rate1} becomes
\be{bound1}
x\,\frac{v}{\gamma}\gsim \left\{ \ba{ll} 1 \hspace{3.1cm}\textrm{Grav. med.} \\ 10^{-2} \left(\frac{\MeV}{m_{3/2}}\right) \qquad \textrm{Gauge med.} \ea \right.
\ee
where we adopted $|K|=0.1$ and $\phi_0=(M_p^{d-3}H/|\lam|)^{1/(d-2)}$. \eq{bound1} shows that for large values of $x$ the condition of thermal equilibrium is fulfilled without assuming any specific form of the distribution. If the Q-balls have a relativistic average velocity, $v\sim 0.1\ldots 1$, then even with $x\sim 1$ (grav.) and $x<1$ (gauge) \eq{bound1} is fulfilled.

One should note that actually the condensate is not formed strictly at the initial times used in \eq{bound1}, which were taken from \cite{kskm1,kskm2,kskm3}, but a little later, as can been seen from simulations in Section 3. For the gravity mediated case the formation time is approximately $t_0\sim 10\ldots 100m_{3/2}^{-1}$ and for the gauge mediated case $t_0\sim 10^5m_{\Phi}^{-1}$. Thus the right-hand side of \eq{bound1} is even smaller by a factor of $0.01\ldots 0.1$.

Another matter is the value of the average charge $\bar{Q}$. If we use a thermal distribution we get from \cite{ejmv} that $\bar{Q}\lsim 10^10$ in the gravity mediated case. This is orders of magnitude smaller than $Q_{\max}$ so that the condition of thermal equilibrium, \eqs{rate1}{bound1}, is fulfilled self-consistently. It was also shown in \cite{ejmv} that even with a low energy-to-charge ratio such as $x\sim 10$ the thermal distribution is a very good approximation to the true distribution. In the gauge mediated case therefore the issue of thermalization of the distribution has not yet been checked in the existing simulations. The existence of thermal distributions has been established in the gravity mediated case for both 2 and 3 spatial dimensions \cite{ejmv,muvi3}.

The average velocity, $v$, is undetermined from these arguments. Energetically the existence of relativistic Q-balls is possible, since even with $v\sim 0.1$ we obtain $\gamma\sim 1.005$, leaving \eq{xappr} essentially unaffected. Only if $v\approx 1$ we obtain $\gamma\gg 1$. Therefore it is natural to expect that Q-balls are relativistic.

We can calculate the absolute minimum of $|x|$ by treating $\dot\phi$, $\phi$, $\dot\theta$ and $\theta$ as independent variables to obtain
\be{xmin}
|x| = \frac{\rho}{m|q|} = \frac{\frac{1}{2}\dot\phi^2 + \frac{1}{2}\dot\theta^2\phi^2 + V(\phi,\theta)}{m|\dot\theta|\phi^2} \geq \sqrt{\frac{2V(\phi,\theta_{\min})}{m^2\phi^2}},
\ee
where we can approximate the potential, $V$, by its mass term alone. In the gravity mediated case this gives $|x|\gsim 1.01$ (for $\phi/M=0.1$ and $K=-0.01$), and in the gauge mediated case $|x|\gsim 10^{-3}\,(10^{-8})$ for $d=4\,(6)$. In practice these are not realized. Because minimum $|x|$ would require a circular orbit. In the next Section we will discuss the realistic values of $x$.

\section{Numerical simulations}
\subsection{Details of the numerical simulations}
We have simulated the evolution of the flat direction starting from the scale of inflation to the time of AD condensate formation. The numerical simulations were done by using routines for ordinary differential equations from the NAG library. We parameterized the AD field $\Phi$ as real and imaginary parts, rescaled the fields and time to be dimensionless and used logarithmic time in order to handle the scale difference between the scale of inflation and AD condensate formation
\be{rescale}
\Phi = \frac{1}{\sqrt{2}} (\phi_1+i\phi_2)\,,\qquad \phi_i = \left(\frac{M_p^{d-3}\,H}{|\lam|}\right)^{1/(d-2)}\,\chi_i\,,\qquad z=\log(mt),
\ee
where $m=m_{3/2}$ in the gravity mediated case and $m=m_{\Phi}$ in the gauge mediated case. With a redefinition of the phase of the field $\Phi$ we put $\theta_A=0$ and $\theta_{\lam}=\pi$ so that the minima are at $b=0$ in \eq{theta1} \ie $d\theta_{\min}=-\theta_a$ for large $H$, when $a\sim 1$, and at $d\theta_{\min}=0$ for small $H$. The initial time is the end of inflation so $z_i=\log(2m/3H_I)$, where $H_I=10^{12}\GeV$ was chosen. Because of the fixed point behaviour \eq{phifix}, the exact value of initial time is not relevant as long as the time, when rotation starts, is much larger than the initial time $t_{rot}\gg t_i$. Then the equations of motion, \eq{eqm1}, become
\bea{eqm4}
\ddot\chi_1 \,+\, C_1\dot\chi_1 \,-\, C_2\chi_1 \,+\,V_m'\chi_1 \,+\, C_3\,(\chi_1^2+\chi_2^2)^{(d-2)/2}\chi_1 \,- & & \nn C_4\,(\chi_1^2+\chi_2^2)^{(d-1)/2}\left[\frac{|A|m_{3/2}}{m}\,e^z\cos((d-1)\theta)+\frac{2|a|}{3}\cos((d-1)\theta+\theta_a)\right] &=& 0, \nn \ddot\chi_2 \,+\, C_1\dot\chi_2 \,-\, C_2\chi_2 \,+\,V_m'\chi_2 \,+\, C_3\,(\chi_1^2+\chi_2^2)^{(d-2)/2}\chi_2 \,+ & & \nn C_4\,(\chi_1^2+\chi_2^2)^{(d-1)/2}\left[\frac{|A|m_{3/2}}{m}\,e^z\sin((d-1)\theta)+\frac{2|a|}{3}\sin((d-1)\theta+\theta_a)\right] &=& 0,
\eea
where
\bea{constants}
C_1 &=& \frac{d-4}{d-2} \nn
C_2 &=& \frac{d-3}{(d-2)^2}+\frac{4c_H}{9} \nn
C_3 &=& \frac{d-1}{9\cdot 2^{d-4}} \nn
C_4 &=& \frac{1}{3\cdot 2^{(d-4)/2}}
\eea
and
\bea{mass}
V_m' \,=\, e^{2z}\left[1-|K|\log\left(\frac{M_p^{d-3}m_{3/2}}{|\lam|}\frac{2}{3}\right)^{2/(d-2)}\frac{\chi_1^2+\chi_2^2}{2M^2}\right]\quad \textrm{(Grav.med.)}, \nn V_m' \,=\, e^{2z}\left[1+\frac{1}{2m_{\Phi}^2}\left(\frac{M_p^{d-3}m_{\Phi}}{|\lam|}\frac{2}{3}e^{-z}\right)^{2/(d-2)}(\chi_1^2+\chi_2^2)\right]^{-1}\quad \textrm{(Gauge med.)}
\eea
and $\theta=\arctan(\chi_2/\chi_1)$, $c_H=|A|=1$, $|a|=0,\,1$. We chose the renormalization scale in the gravity mediated case to be $M=\left(M_p^{d-3}m_{3/2}/|\lam|\right)^{1/(d-2)}$ as was done in \cite{ejm}. However, we have also checked the results for larger values of $M$ that have been used in simulations all the way up to Planck scale \cite{kskm2,kskm3,ejmv,muvi3}. We comment on the differences when discussing the results. The initial conditions \eqs{theta1}{phi1} in these variables read
\bea{init}
\sqrt{\chi_1^2+\chi_2^2} &\approx & \sqrt{2}\left(\frac{|a|+\sqrt{|a|^2+4(d-1)c_H}}{2(d-1)}\right)^{1/(d-2)} \nn \theta_i &=& \left\{\ba{ll} \in [-\frac{\pi}{d},\frac{\pi}{d}],\qquad a\sim 0 \qquad \textrm{(D-term)} \\ -\frac{\theta_a}{d},\qquad a\sim 1,\,\theta_a\in [-\pi,\pi] \qquad \textrm{(F-term)} \ea \right.
\eea
We have restricted the initial phase, $\theta_i$ to this sector since changing the initial phase by $\theta\rightarrow\theta+\frac{2\pi}{d}n$ gives the same result for charge and energy of the condensate. The solution of $\Phi$ in another sector of initial conditions is achieved by a phase shift of $2\pi n/d$ from the sector prescribed.

\subsection{Outline of the numerical results}
\fig{plotreim} shows examples of the rotation of the AD condensate. In the gravity mediated case, \fig{plotreim}(a), we see that the orbit is a spiraling ellipse and in the gauge mediated case, \fig{plotreim}(b), a precessing trefoil. From \fig{plotreim} one can see that there is a twist on the orbit much before the rotation starts properly. This is the time of the phase transition, when the AD field $\Phi$ starts to rotate in the pit of the symmetry breaking minimum. The rotation begins when the symmetry breaking minimum is the vacuum, ends when it has become a false vacuum and twists when the false vacuum has completely vanished forming a kink on the orbit. It is possible to produce a condensate through a second order phase transition, too, but the charge in that case would be small. It should also be noted that in the gravity mediated case condensate formation starts when $c_HH^2\sim m_{3/2}^2$ for all values of $d$, $A$ and $a$. In the gauge mediated case the condensate formation starts at $c_HH^2\sim m_{\Phi}^4/|\Phi|^2$, so that the formation happens earlier if the mass, $m_{\Phi}$, is increased, as can be seen from the different positions of the kink in \fig{plotreim}(b).

\bfig
\leavevmode
\centering
\vspace*{4cm}
\includegraphics{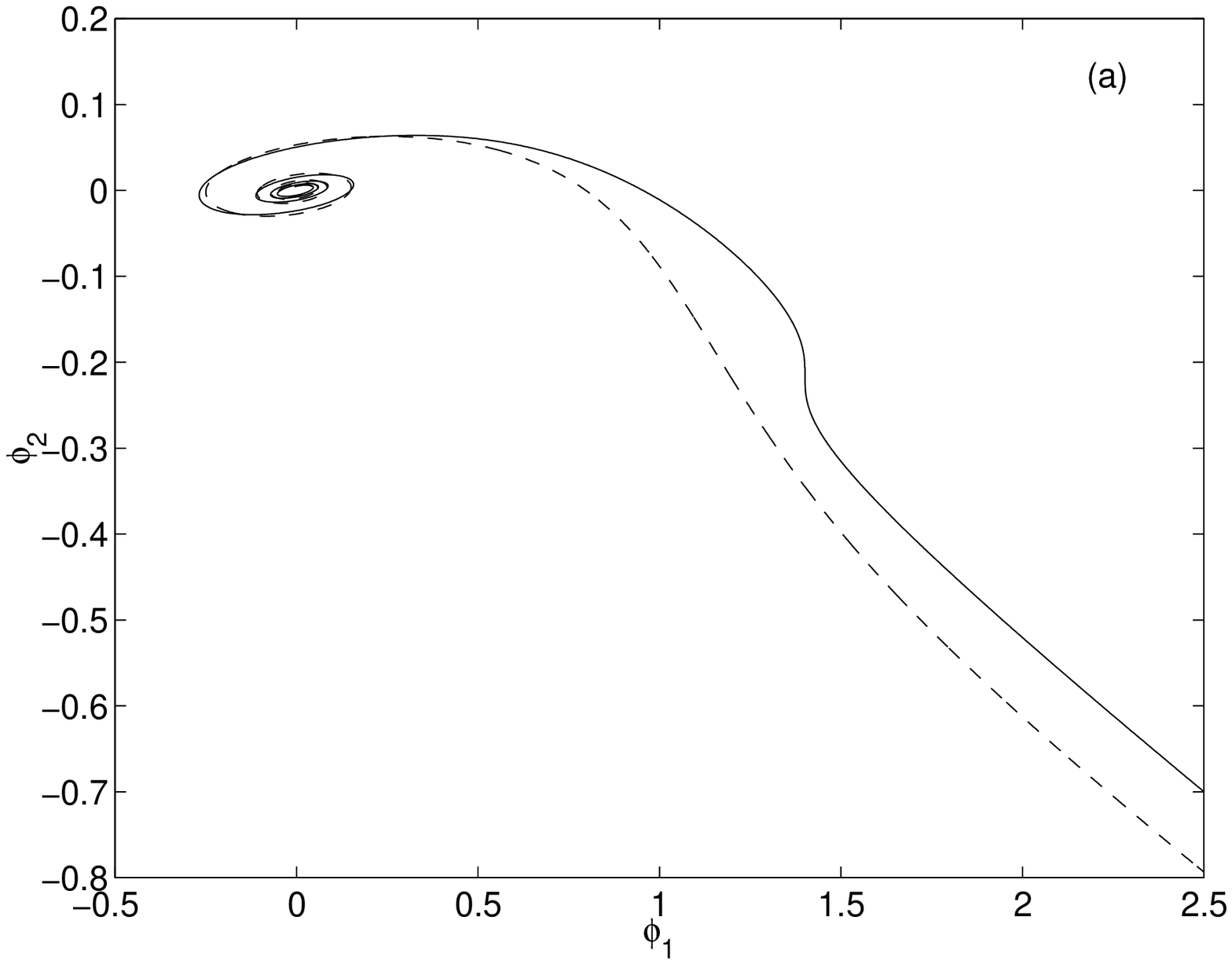}
\includegraphics{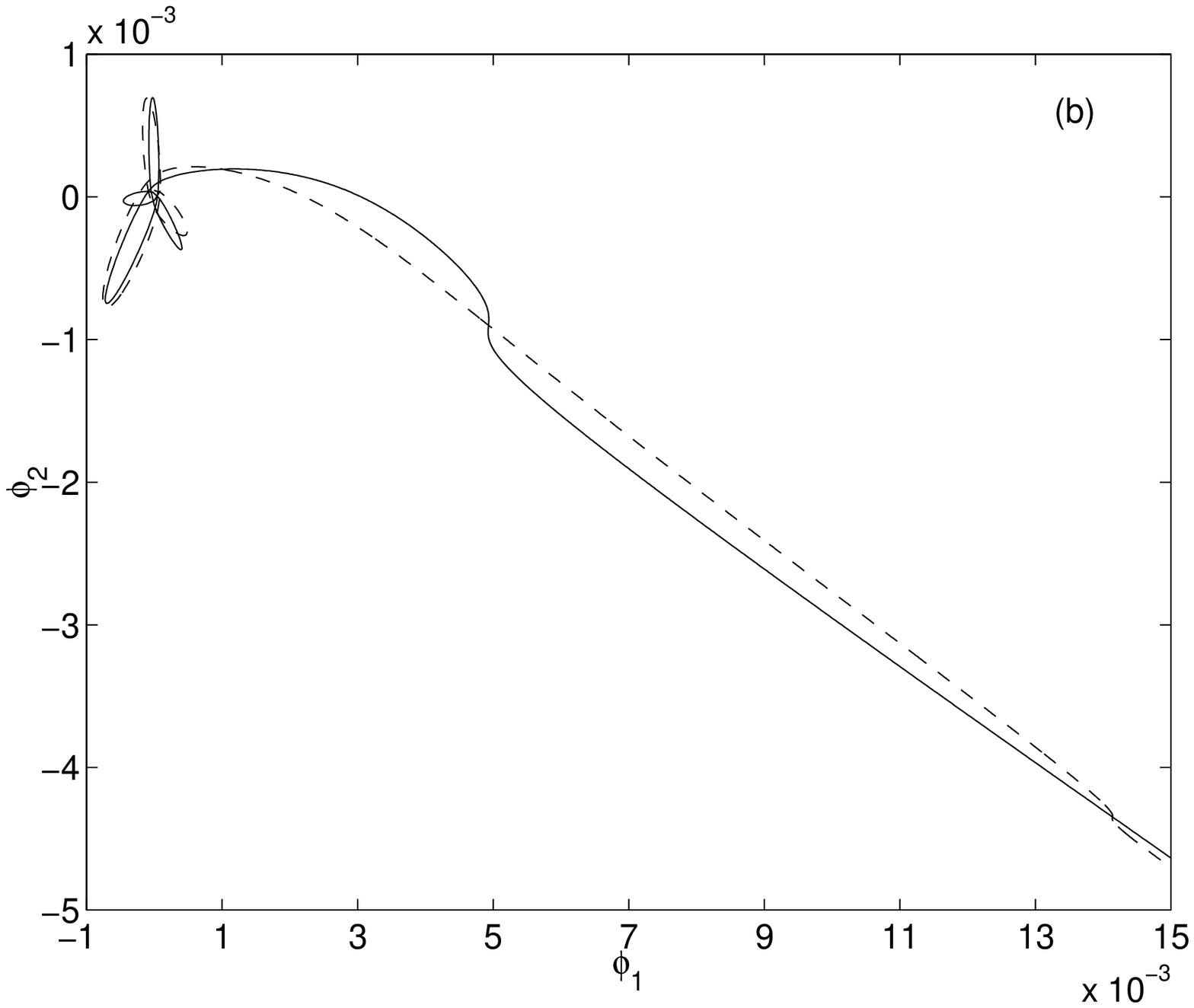}
\caption{\small Affleck-Dine condensate formation (a) gravity mediated case with $d=4$ (solid) and $d=6$ (dashed) and (b) gauge mediated case with $d=4$, $m_{\Phi}=1\TeV$ (solid) and $m_{\Phi}=10\TeV$ (dashed) with initial condition $\theta_i=-\pi/10$ and $a=0$ in both figures.}\label{plotreim}
\efig

The quantities of interest are the charge density $q$, the energy-to-charge ratio $x$, the pressure-to-energy ratio $w$, the ellipticity of the orbit $\eps$ and the decay of the AD field amplitude $\phi$. The charge density contributes to the baryon-to-entropy ratio and is depicted in Figs. \ref{plotchagr} and \ref{plotchaga}. The energy-to-charge ratio $x$, which gives the relative amount of Q-balls, anti-Q-balls and their kinetic energies, \eq{xappr}, is shown in Figs. \ref{plotenchagr} and \ref{plotenchaga}. The pressure-to-energy ratio $w$, which gives the equation of state and indicates whether the condensate is unstable or not, is shown in Figs. \ref{plotpreengr} and \ref{plotpreenga}, where we also show the ellipticity of the orbit $\eps$ (the calculation of the average $w$ is technically somewhat problematic, because of the rapid oscillations). The of the AD field amplitude $\phi$, to verify the proposed behaviour \eq{fix}, is shown in \fig{plotphigr}.

We display the results for both D- and F-term inflation in the gravity and gauge mediated cases: we adopt dimensions $d=4,\,5,\,6,\,7$ (gravity) and $d=4,\,6$ (gauge). The gravitino mass in the gauge mediated case is chosen to be $m_{3/2}=10^{-5}m_{\Phi}$ ($d=4$) and $m_{3/2}=10^{-9}m_{\Phi}$ ($d=6$). The variation in $m_{3/2}$ only results in variation in the charge density and therefore affects $x$ inversely. In the gravity mediated case the oscillation of $x$ asymptotes at around $t=100m_{3/2}^{-1}$. In the gauge mediated case there is no asymptotic behaviour but $x$ continues to increase steadily. Therefore in the gauge mediated case the quantities in the Figures are presented for different times.

We have summarized the values of $q$, $x$, $w$ and $\eps$ in Table 1 (gravity) and Table 2 (gauge).

\subsubsection{Gravity mediated case, D-term}
In this case $a=0$ and the mass term is given by \eq{mgrav}. Initially the phase, $\theta_i$, is random and $\phi$ is at the instantaneous minimum \eq{phi1}. The parameter values adopted are $M=(M_p^{d-3}m_{3/2}/|\lam|)^{1/(d-2)}$; $|A|=c_H=1$; $K=-0.01$ (and $K=-0.1$); and $d=4,\,5,\,6,\,7$.

\bfig
\leavevmode
\centering
\vspace*{3cm}
\includegraphics{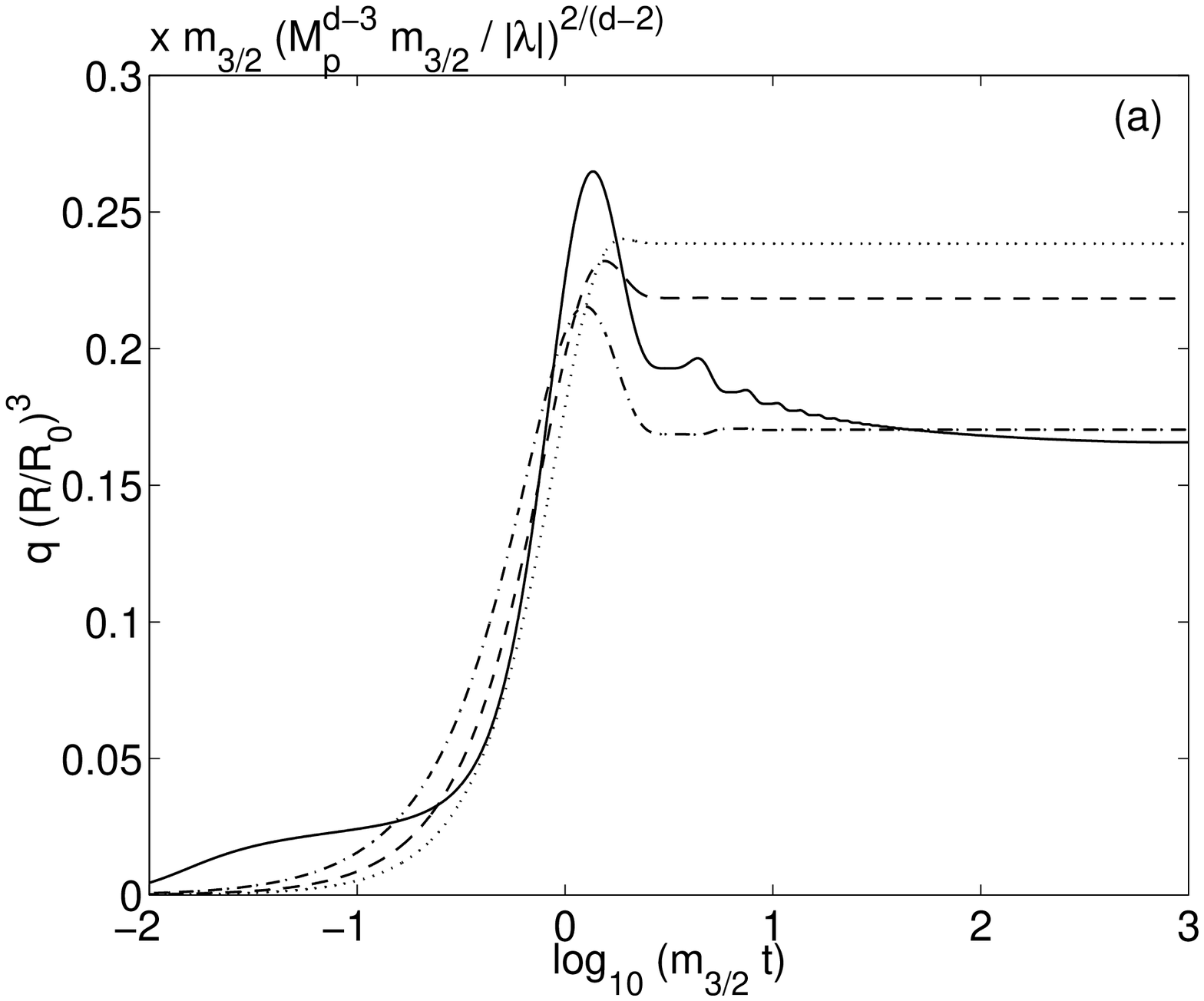}
\includegraphics{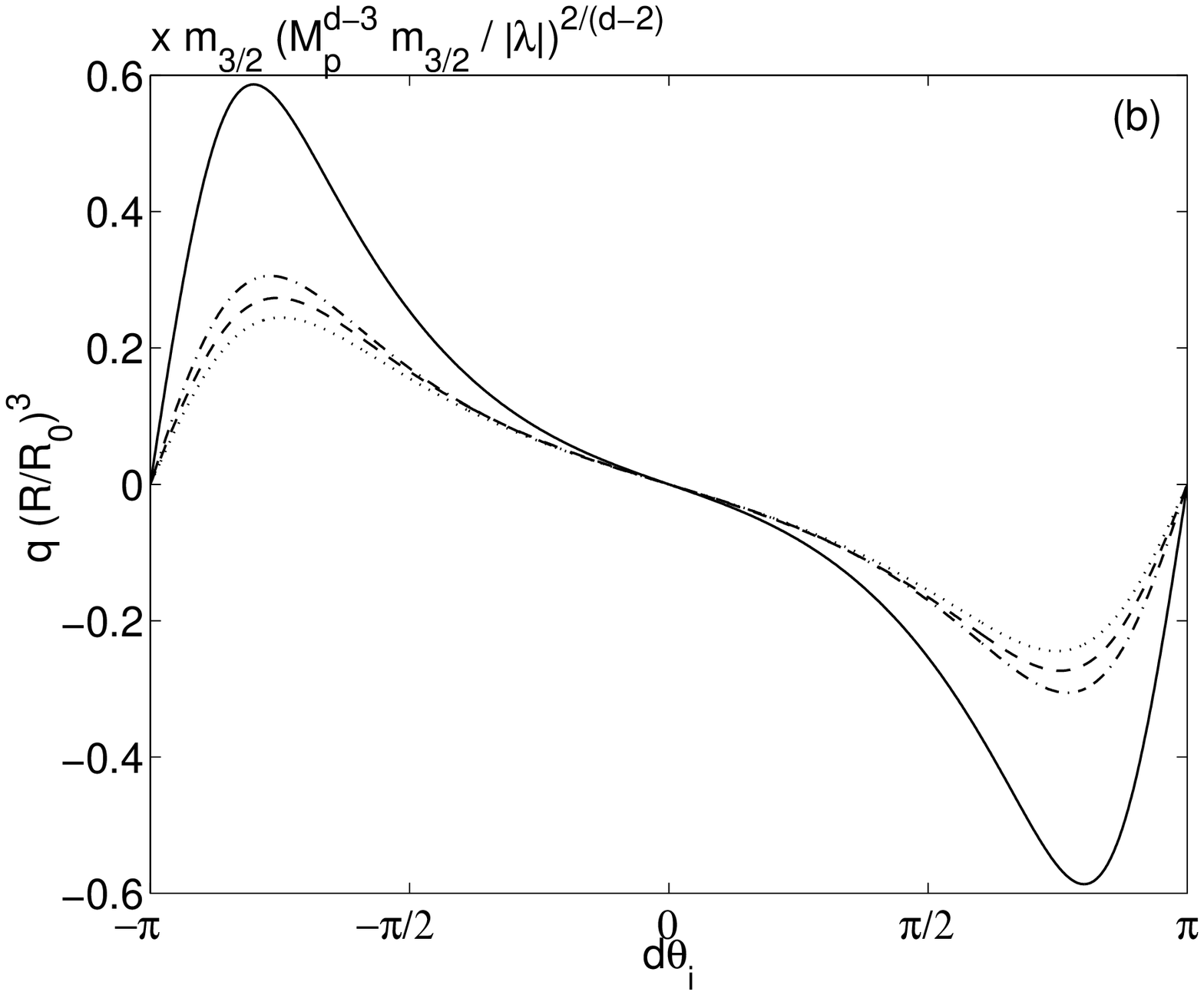}
\includegraphics{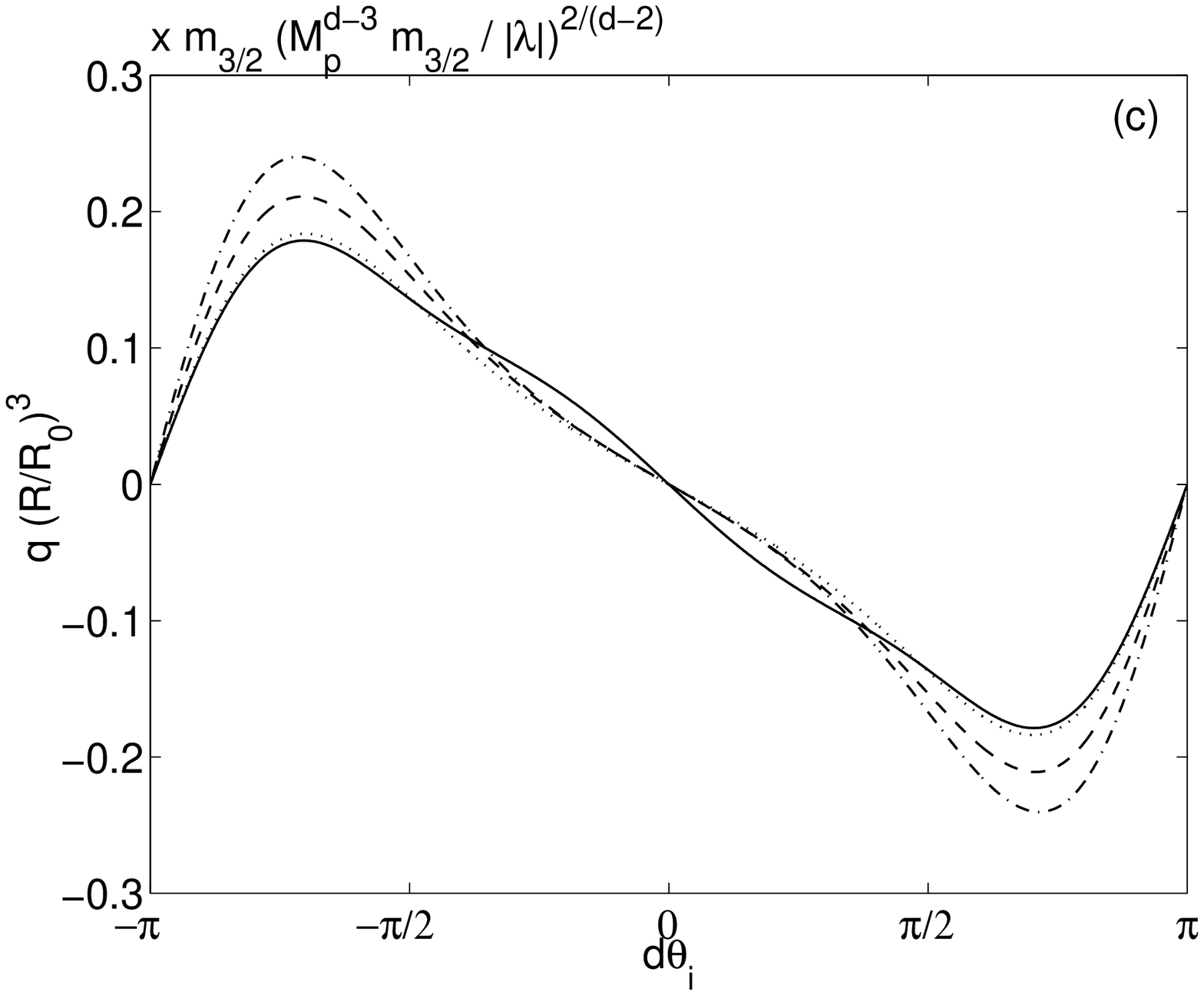}
\caption{\small Charge density in the co-moving volume in the gravity mediated case vs. (a) time in logarithmic units of time, (b) D-term ($a=0$) and (c) F-term ($a=1$) case with $d=4,5,6,7$ (solid, dash-dot, dashed and dotted lines), $K=-0.01$ and $t=100m_{3/2}^{-1}$.}\label{plotchagr}
\efig

In \fig{plotchagr}(a) we show the time evolution of the co-moving charge density, $q(R/R_0)^3$. One can see that it becomes a constant when $H\sim 0.1\ldots 0.01m_{3/2}$. In \fig{plotchagr}(b) the charge density in the co-moving volume is plotted against the different initial conditions. It can be seen that the charge density is positive for $-\pi<d\theta_i<0$ and negative for $0<d\theta_i<\pi$. However, this behaviour depends on the values of $c_H$ and $A$. For instance, here we have chosen $c_H=1$ whereas in \cite{drt} the choice $c_H=9/4$ was made resulting in a negative charge density for some values $d\theta_i<0$. This comes about because the AD-field $\Phi$ manages to rotate around the symmetry breaking minimum before it vanishes, thereby gaining phase motion in the opposite direction (we have verified this behaviour). Therefore the sign of the charge cannot be given generically for any range of initial conditions. Only the magnitude of the charge density can be given and is about the order depicted in \fig{plotchagr}(b). Typically the charge density lies in the range $q(R/R_0)^3\sim 0.1\ldots 0.6m_{3/2}(M_p^{d-3}m_{3/2}/|\lam|)^{2/(d-2)}$. This results in a charge-to-entropy ratio $q/s\sim 0.01\ldots 0.1\,(T_R/m_{3/2})\,(m_{3/2}/|\lam|M_p)^{2/(d-2)}$, where $T_R$ is the reheating temperature. Here we have assumed that the entropy density $s\approx 4\rho_I/T_R$, where the inflaton energy density is $\rho_I\approx 3M_p^2H^2$. For $d=4$ and $m_{3/2}=100\GeV$ this would require $T_R\sim 10^9\ldots 10^{10}\GeV$ if the baryon-to-entropy ratio, $n_B/s\sim 10^{-10}$, is to be explained through AD mechanism. For $d=6$ we would obtain $T_R\sim 10\ldots 100\GeV$. These are of the same order as the results of \cite{drt} such that the radiative correction does not produce any significant variation.

\bfig
\leavevmode
\centering
\vspace*{3cm}
\includegraphics{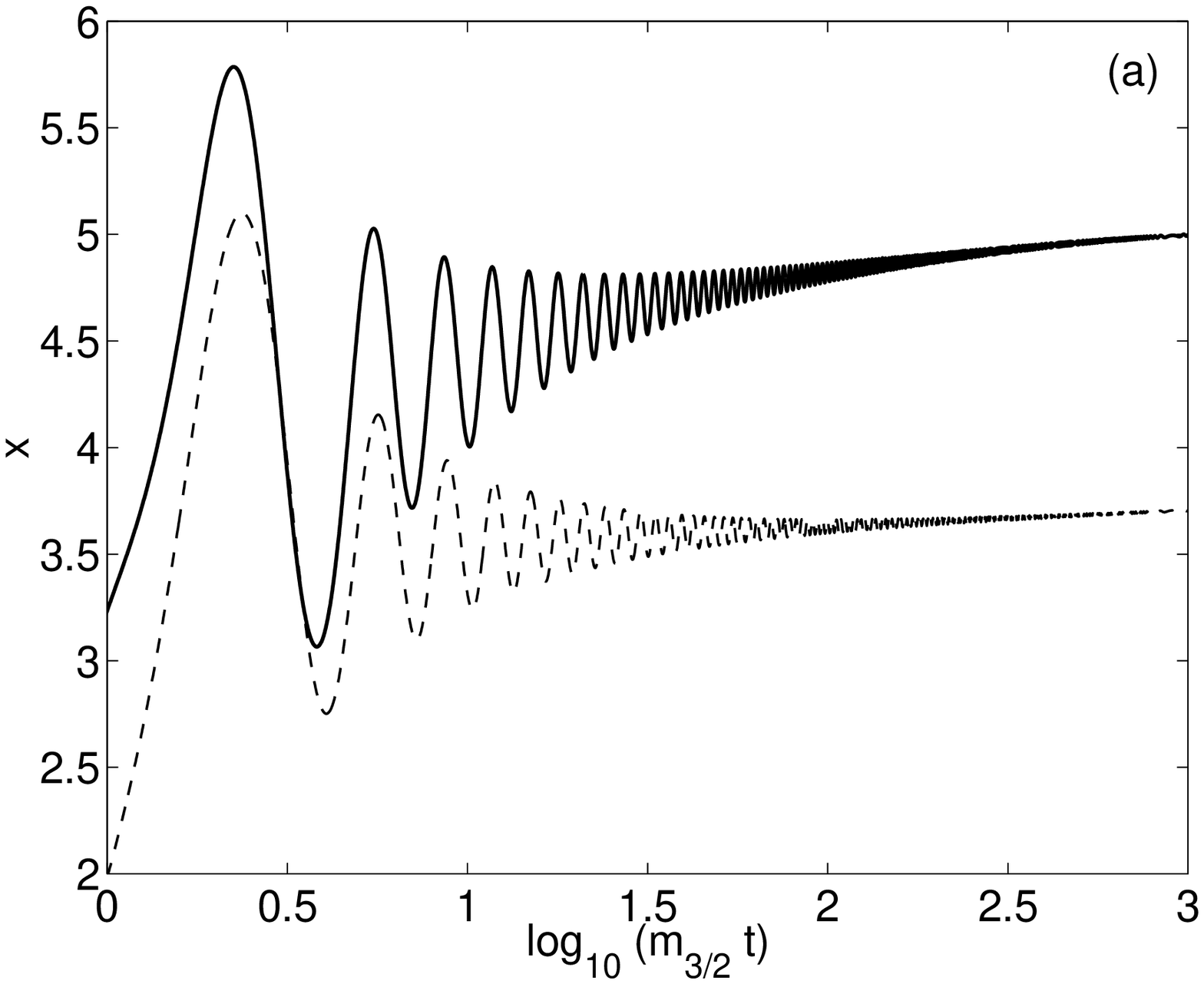}
\includegraphics{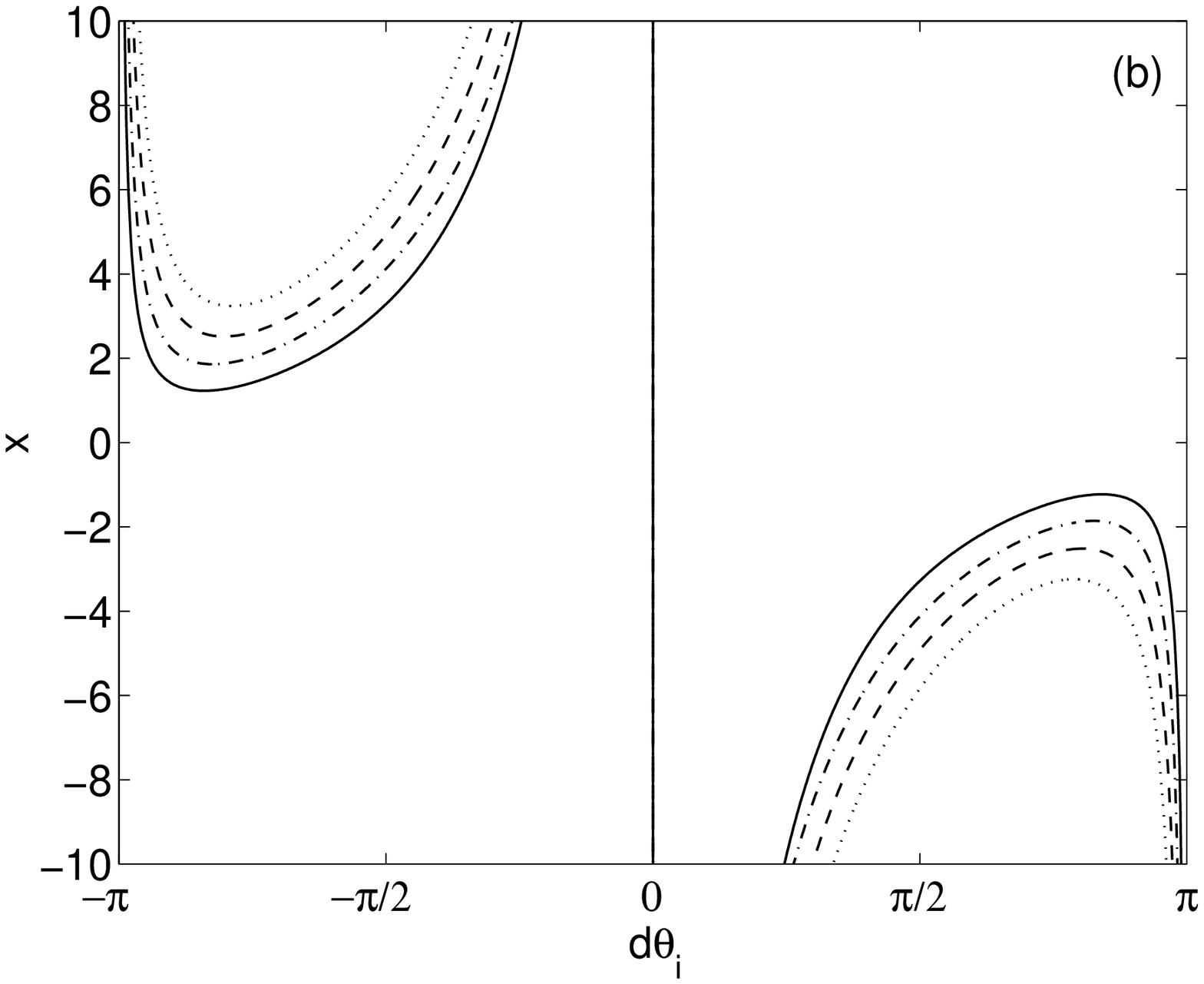}
\includegraphics{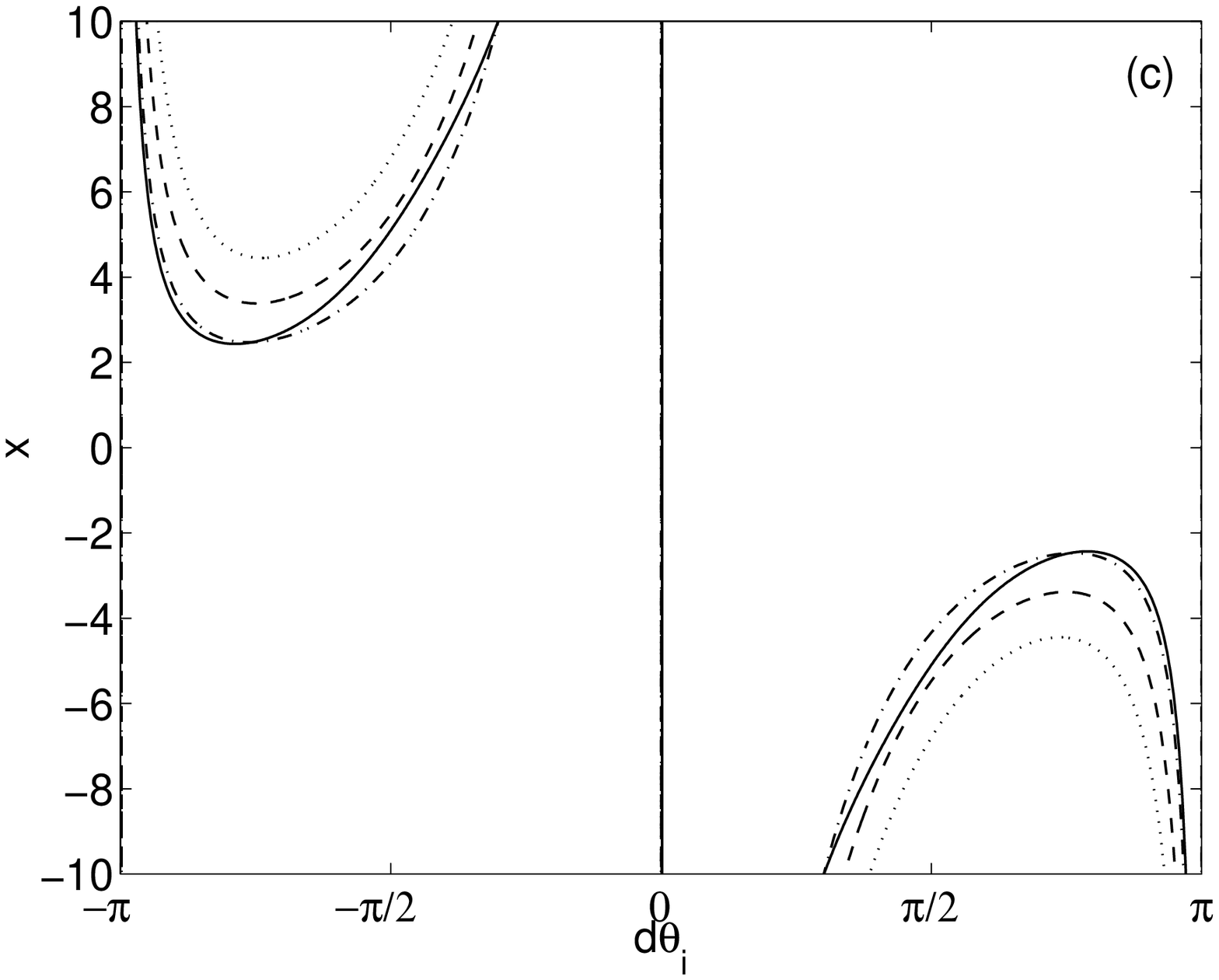}
\caption{\small Energy-to-charge ratio, $x$, in the gravity mediated case vs. (a) time in logarithmic units for $d=4,\,6$, (b) D-term ($a=0$) and (c) F-term ($a=1$) case with $d=4,5,6,7$ (solid, dash-dot, dashed and dotted lines), $K=-0.01$ and $t=100m_{3/2}^{-1}$.}\label{plotenchagr}
\efig

In \fig{plotenchagr}(a) we depict the time development of the energy-to-charge ratio $x$. One can see that $x$ oscillates strongly with a decaying amplitude and the overall behaviour is that $|x|$ increases. At $t=100m_{3/2}^{-1}$ the oscillation amplitude has dampened enough such that $x$ is approximately a constant. In \fig{plotenchagr}(b) we plot $x$ against the different initial conditions. There one can see that as $d$ increases, $|x|$ also increases. Roughly one third of the total range of the initial conditions were found to have $|x|>10$ while two thirds have $1.1<|x|<10$ (lower limit is the minimum value for $d=4$). Therefore more simulations for $|x|<10$ should be made to establish whether the thermal distribution is valid or not. If $|K|$ or $M$ are increased from the values used here, $|x|$ becomes slightly larger.

The oscillation of pressure-to-energy density ratio, $w$, is plotted in \fig{plotpreengr}(a), which shows that the oscillation of $x$ corresponds to oscillation of pressure from positive to negative. The average pressure is slightly to the negative side, resulting in an increasing $x$. We have calculated the average $w$ in \fig{plotpreengr}(b) at $t\sim 100m_{3/2}^{-1}$ with different initial conditions. In \fig{plotpreengr}(c) the ellipticity of the orbit, $\eps$, is plotted to show that $w$ is more negative if $\eps$ is small. However, it should be noted that $w$ achieves values which are more negative than the absolute lower bound coming from pure oscillation, see discussion after \eq{w2}. Therefore we also show the average $w$ at $t=300m_{3/2}^{-1}$ in the same figure. This behaviour might be due to $\partial V/\partial t$ in \eq{cont}, whose contribution decreases as time evolves. It is also due to the rapid oscillation of $w$ that there are numerical inaccuracies in our calculations. However, we believe that this does not affect our main conclusions.

\bfig
\leavevmode
\centering
\vspace*{4cm}
\includegraphics{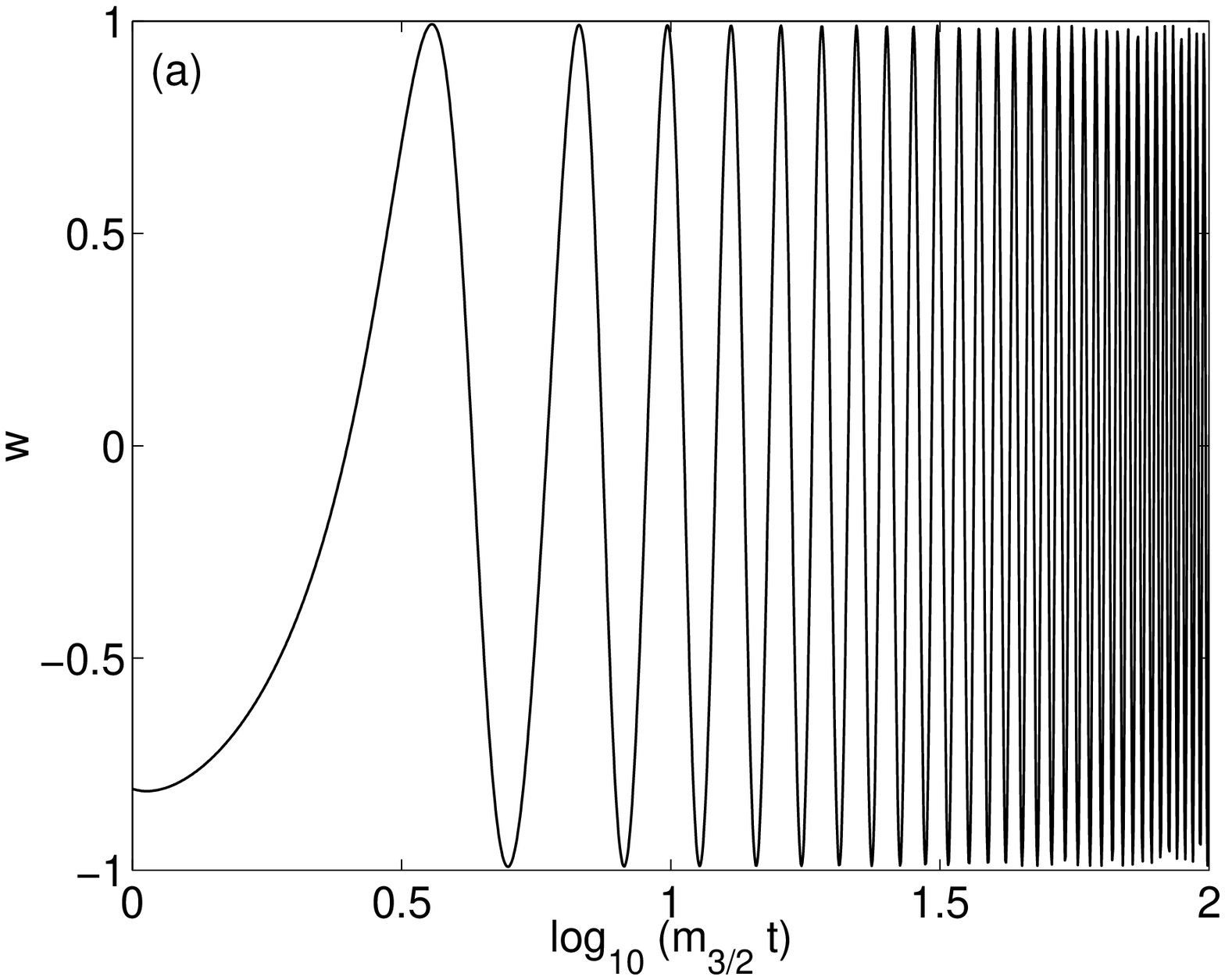}
\includegraphics{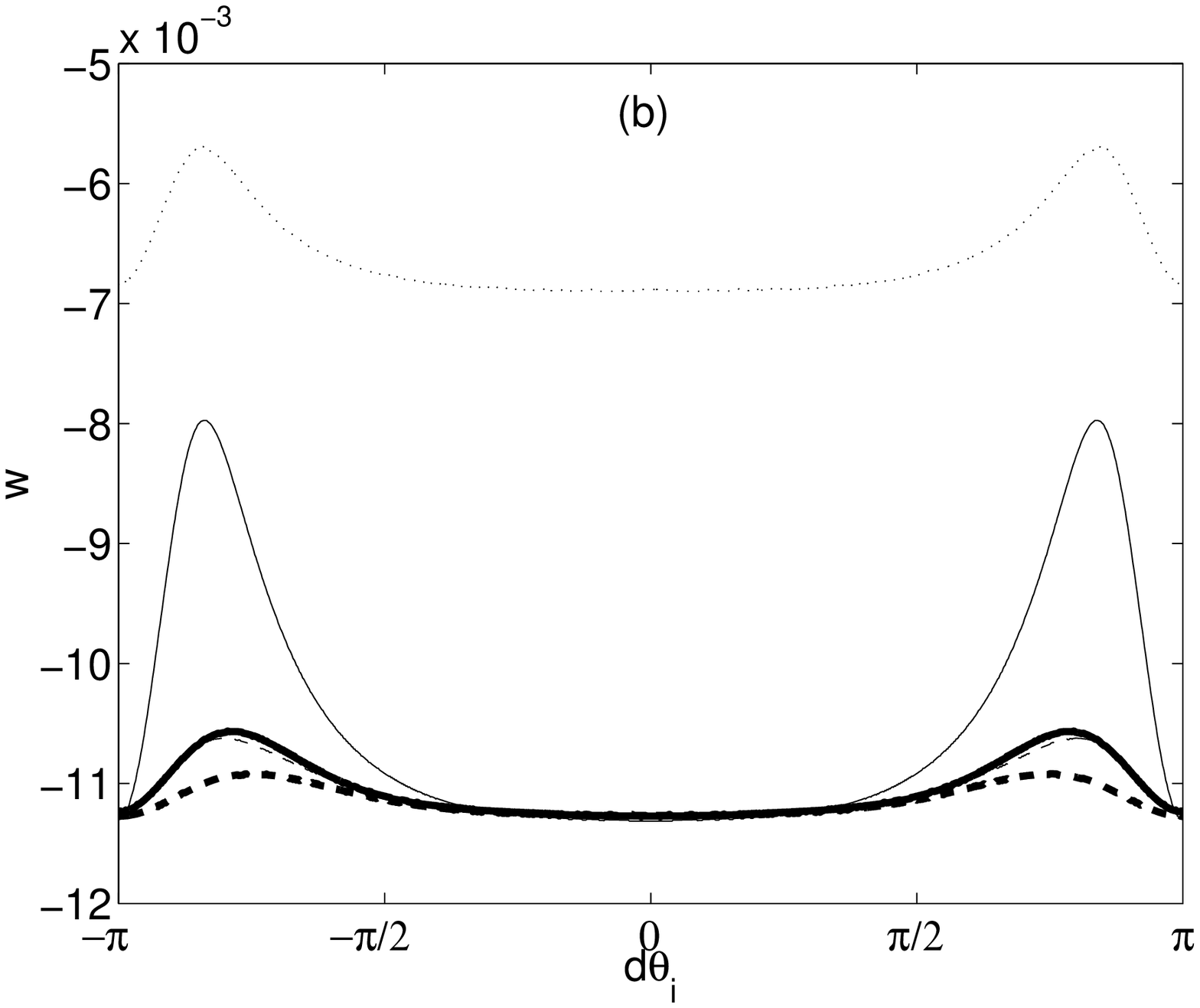}
\includegraphics{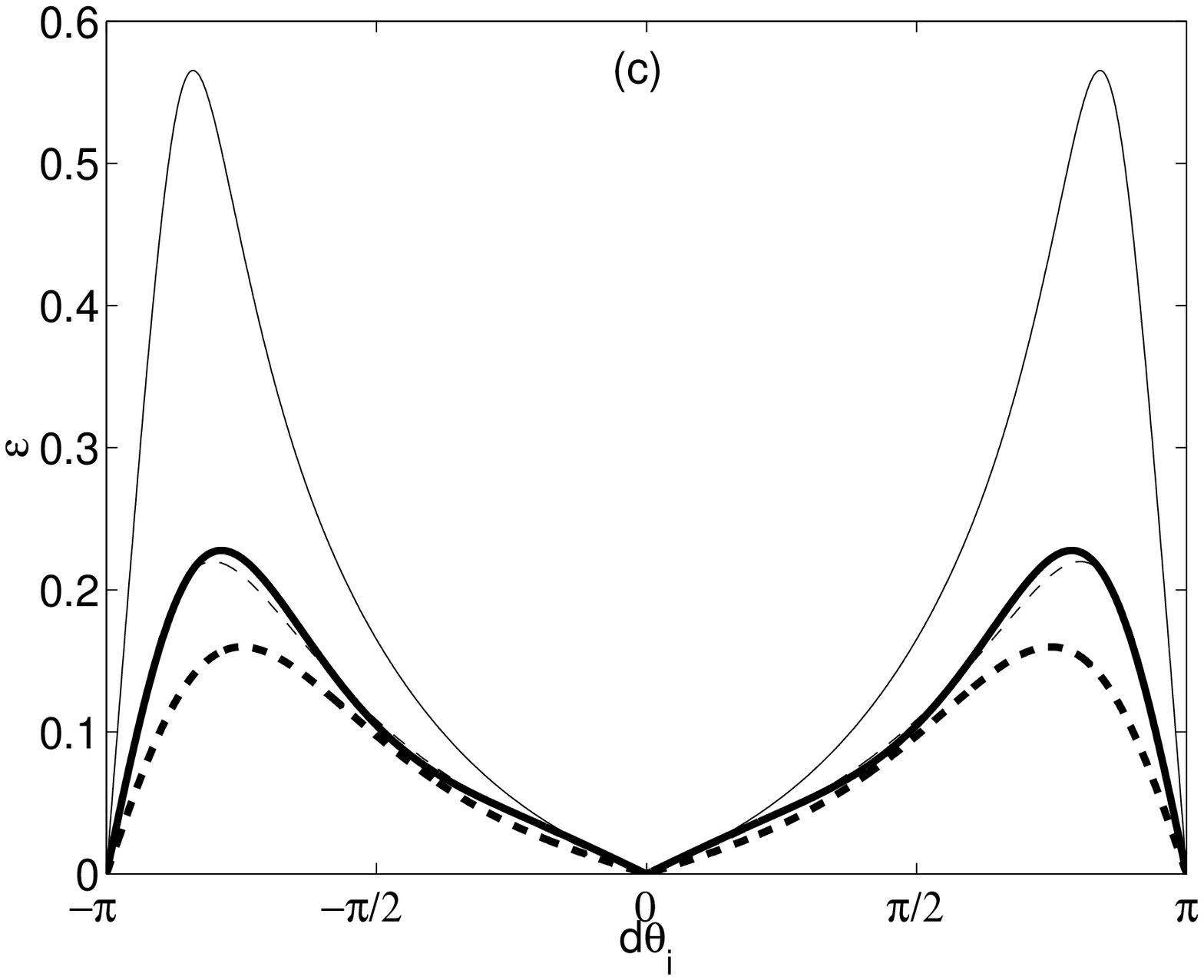}
\caption{\small Pressure-to-energy density ratio, $w$, in the gravity mediated case vs. (a) time in logarithmic units for $d=4$ and (b) different initial conditions for $d=4,\,6$ and (c) ellipticity, $\eps$, vs. initial conditions. $d=4$ (thin lines), $d=6$ (thick lines), D-term (solid), F-term (dashed) with  $K=-0.01$ and $t=100m_{3/2}^{-1}$. In plot (b) there is plotted $w$ at $t=300m_{3/2}^{-1}$ with dotted lines for $d=4$ D-term case.}\label{plotpreengr}
\efig

From \fig{plotphigr}(a) one can see that the amplitude, $\phi$, decays as $(R/R_0)^{-3/(2+K)}$ where $K=-0.01$. We checked that for $K=-0.1$ the decay of $\phi$ tends to slow down. This is due to the fact that for $K=-0.1$ the approximation of the mass term as a polynomial breaks down and in effect gives rise to smaller $|K|$. However, the effect is not significant unless $M$ very large.

\subsubsection{Gravity mediated case, F-term}
In this case $|a|=1$ and the initial conditions are given by \eq{init}. There is however no practical difference with respect to the D-term case, only minor quantitative changes. In \figs{plotchagr}{plotenchagr}(c) we have plotted the charge density in the co-moving volume and $x$ to compare them with the D-term case. Charge densities are approximately similar for $d=5,\,6,\,7$ but for $d=4$ there is a difference. This is probably due to the nature of the fixed point \eq{phifix}: for $d=4$ it is marginal whereas for $d>4$ it is attractive \cite{drt}. Now $|x|$ is slightly larger than in the D-term case, the minimum being $|x|\gsim 2$. The orbits are also slightly more elliptic, as can be seen in \fig{plotpreengr}(c) resulting in more negative pressure, see \fig{plotpreengr}(b).

\bc {\bf Table 1. Gravity mediated case} \ec
\btab{|c|c|c|}
\hline
$d$ & $4$ & $6$ \\
\hline
$q(R/R_0)^3\,/\,m_{3/2}(M_p^{d-3}m_{3/2}/|\lam|)^{2/(d-2)}$ & $\sim 0.1\ldots 0.6$ & $\sim 0.1\ldots 0.3$ \\
\hline
$|x|$ & $>1.1$ & $>2.5$ \\
\hline
$w$ & $\sim -(0.05\ldots 1)|K|/2$ & $\sim -(0.2\ldots 1)|K|/2$ \\
\hline
$\eps$ & $<0.5$ & $<0.2$ \\
\hline
\etab

\subsubsection{Gauge mediated case, D-term}
In the gauge mediated case the mass term is given by \eq{mgauge}. Now $a=0$. We have checked the cases $d=4,\,6$ with $m_{3/2}=10^{-5},\,10^{-9}m_{\Phi}$, $|A|=c_H=1$ and $m_{\Phi}=1,\,10,\,100\TeV$. Here we have chosen $m_{3/2}$ such that a maximal amount of charge, and therefore the minimum value of $|x|$ is produced. If $m_{3/2}$ were larger, we would have to include the mass term of the gravity mediated case and the behaviour would effectively be that of the gravity mediated type already discussed.

\bfig
\leavevmode
\centering
\vspace*{3cm}
\includegraphics{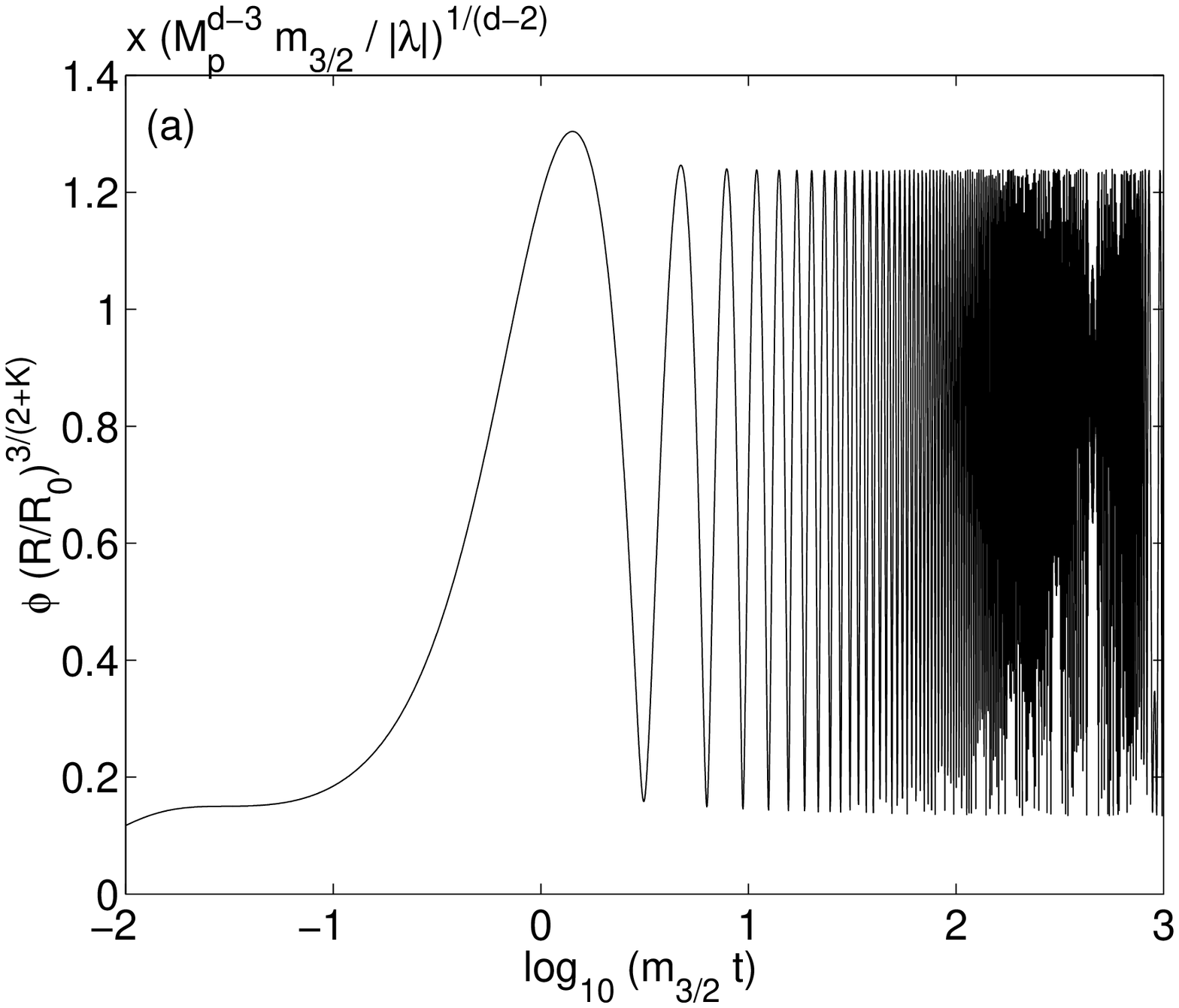}
\includegraphics{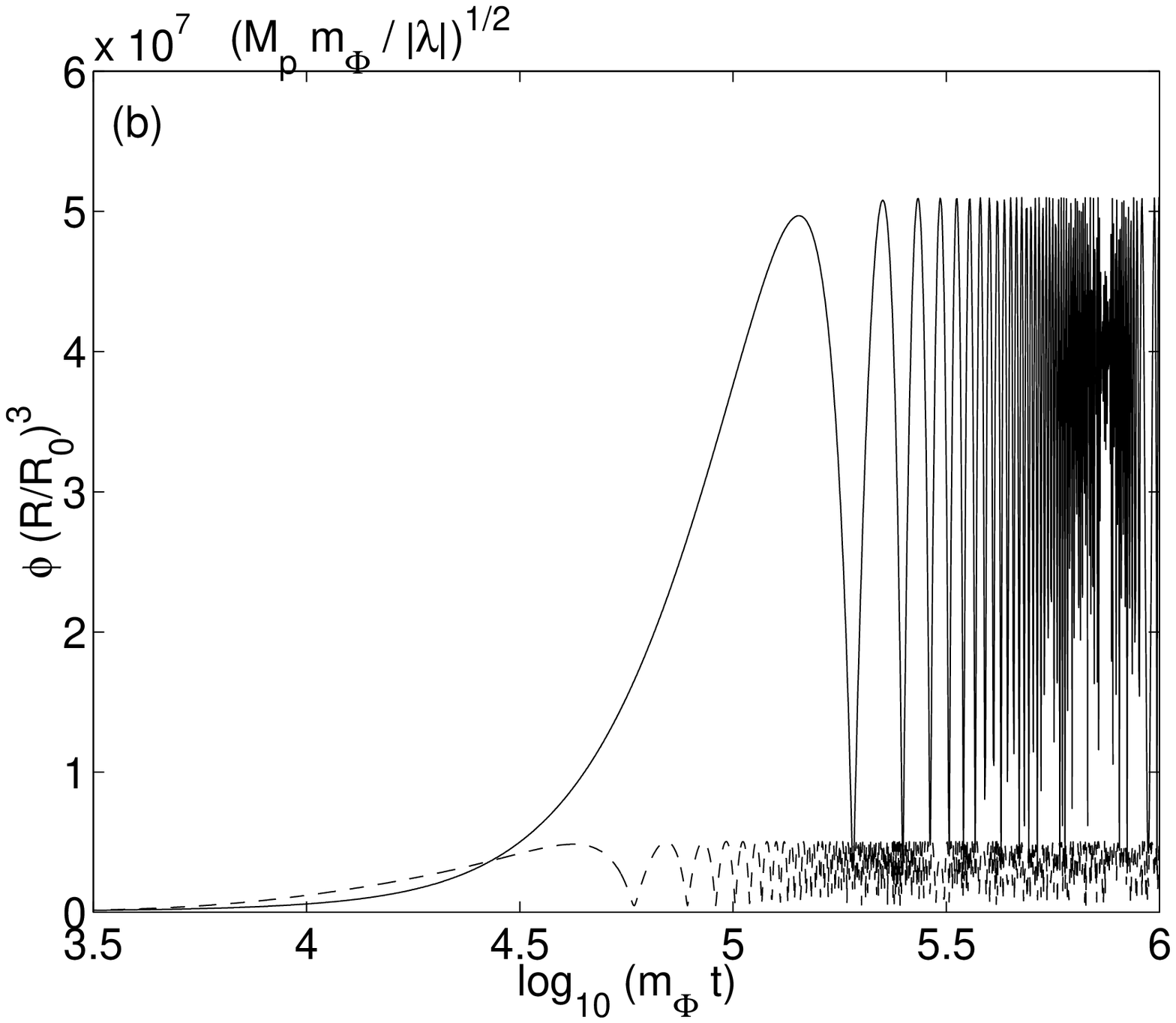}
\includegraphics{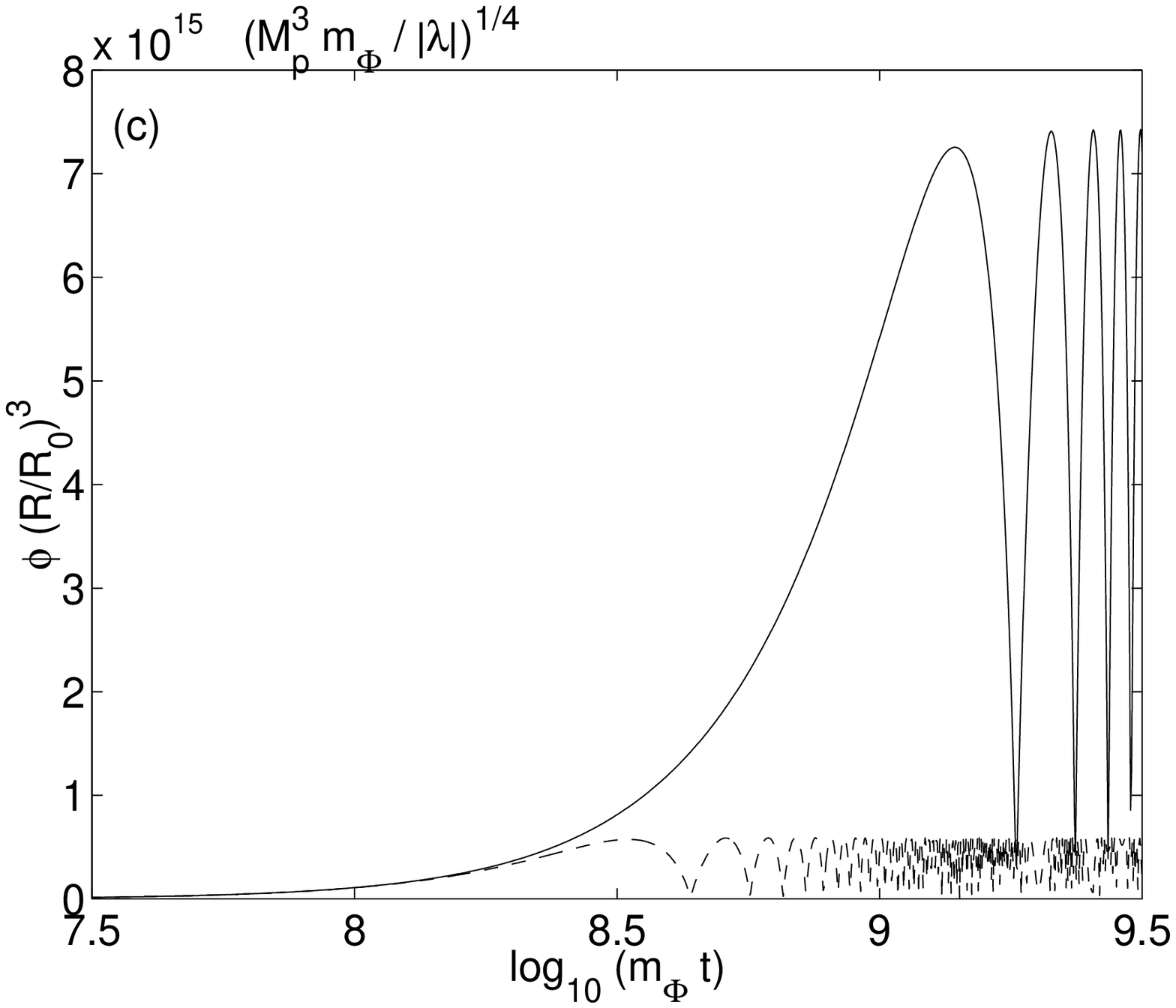}
\caption{\small Time development of the field amplitude (a) $\phi(R/R_0)^{3/(2+K)}$ in the gravity mediated case for $d=4$ and $K=-0.01$; gauge mediated case $\phi(R/R_0)^3$ for (b) $d=4$ and (c) $d=6$.}\label{plotphigr}
\efig

In \fig{plotchaga}(a) the time development of the charge density in the co-moving volume, $q(R/R_0)^3$, is shown for $d=4$. One can see that it reaches its asymptote earlier for large $m_{\Phi}$. The charge in dimensionless units is different for different $m_{\Phi}$. This is due to the fact that one cannot simultaneously remove the dependence on $m_{\Phi}$ and $m_{3/2}$. The resulting charge density in the co-moving volume is $q(R/R_0)^3\sim 0.1M_pm_{\Phi}^2/|\lam|$ for $d=4$ and $q(R/R_0)^3\sim 1000m_{\Phi}(M_p^3m_{\Phi}/|\lam|)^{1/2}$ for $d=6$. This results into charge-to-entropy ratios $q/s\sim 0.1 T_R/M_p$ for $d=4$ and $q/s\sim 100T_RM_p^{-1/2}m_{\Phi}^{-1/2}$ for $d=6$.

\bfig
\leavevmode
\centering
\vspace*{3cm}
\includegraphics{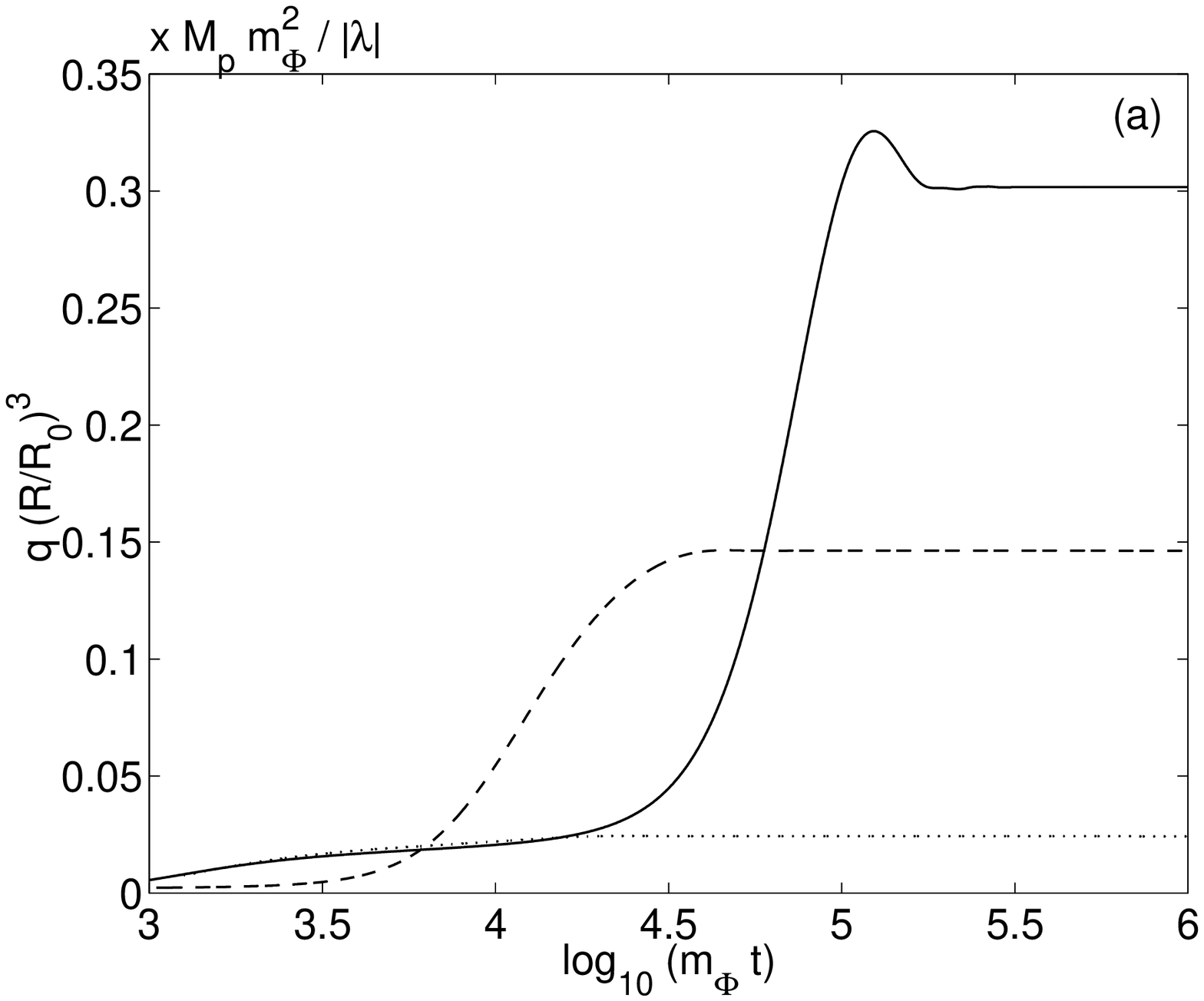}
\includegraphics{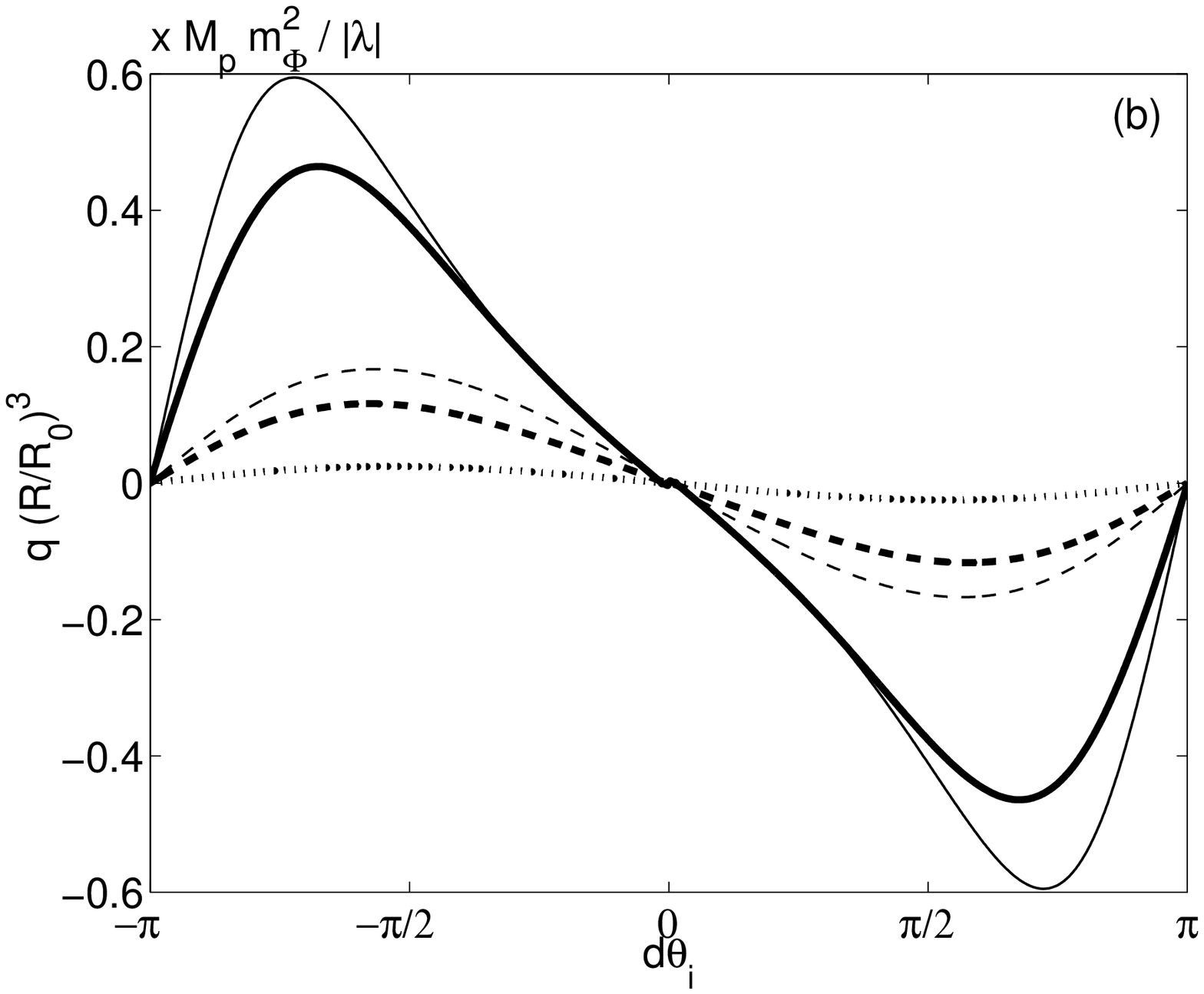}
\includegraphics{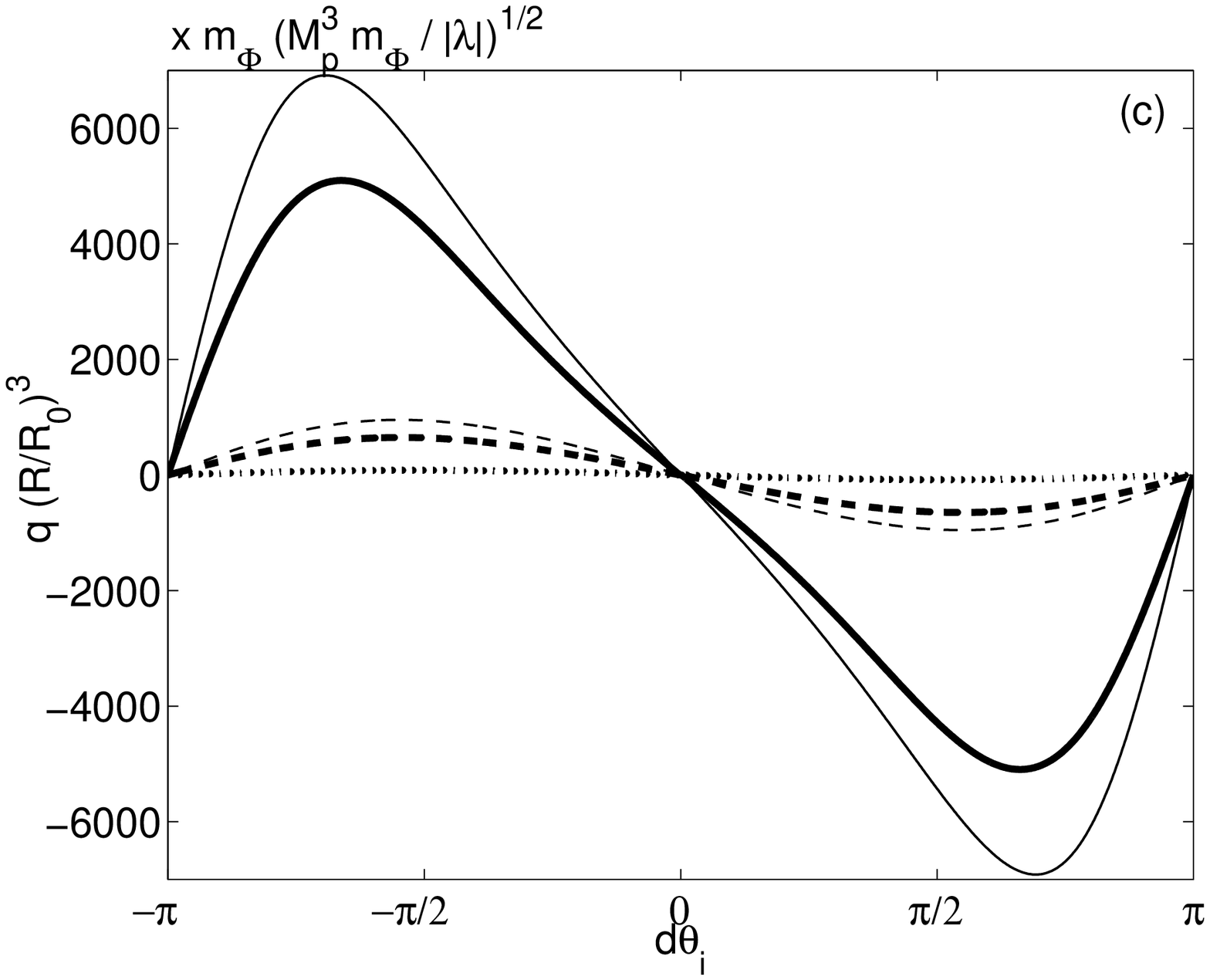}
\caption{\small Charge density in the co-moving volume in the gauge mediated case vs. (a) time in logarithmic units of time $d=4$, (b) $d=4$ and (c) $d=6$ with D-term (thin lines, $a=0$) and F-term (thick lines, $a=1$) and $m_{\Phi}=1,\,10,\,100\TeV$ (solid, dashed, dotted lines) at $t=4\cdot 10^5m_{\Phi}^{-1},\,10^5m_{\Phi}^{-1},\,4\cdot 10^4m_{\Phi}^{-1}$ ($d=4$) and $t=4\cdot 10^9,\,10^9,\,4\cdot 10^8m_{\Phi}^{-1}$ ($d=6$).}\label{plotchaga}
\efig

In \fig{plotenchaga}(a) the time development of energy-to-charge ratio, $x$, is shown for $d=4$. One can see that there is no oscillation in the gauge mediated case in contrast to the gravity mediated case. $x$ increases faster, which is due to the behaviour $\phi\sim R^{-3}$, see \fig{plotphigr}(b), which causes the rotation velocity $\dot\theta\sim R^3$ to increase rapidly. In Figs. \ref{plotenchaga}(b) and (c) we show the $d=4$ and $d=6$ cases. One can see that for $d=4$ one obtains $|x|\gsim 10^{-2}$ whereas $d=6$ results in $|x|\gsim 10^{-5}$ (for $m_{\Phi}=1\TeV$, which gives the lowest bounds). If $m_{3/2}$ decreases, the charge density becomes smaller, thus making $|x|$, which is inversely proportional to charge density, larger. In a realistic case one should take the gravity mediated mass term, \eq{mgrav}, together with the gauge mediated mass term, \eq{mgauge}, for these borderline cases. This would increase $|x|$, so that the true minimum values are likely to be an order of magnitude larger. One should also note that in order for the AD condensate to form via the gauge mediation mechanism one requires $m_{3/2}<0.1\ldots 10\GeV$ for $d=4$ or $m_{3/2}<1\ldots 100\eV$ for $d=6$. For $d=4$ condensate formation is quite possible, but for $d=6$ it is likely that a gravity mediated mass term must be included. Then it would appear that the AD condensate forms through gravity mediation, but the Q-balls via gauge mediated mechanism or via a mixture of these \cite{kskm4}.

\bfig
\leavevmode
\centering
\vspace*{4cm}
\includegraphics{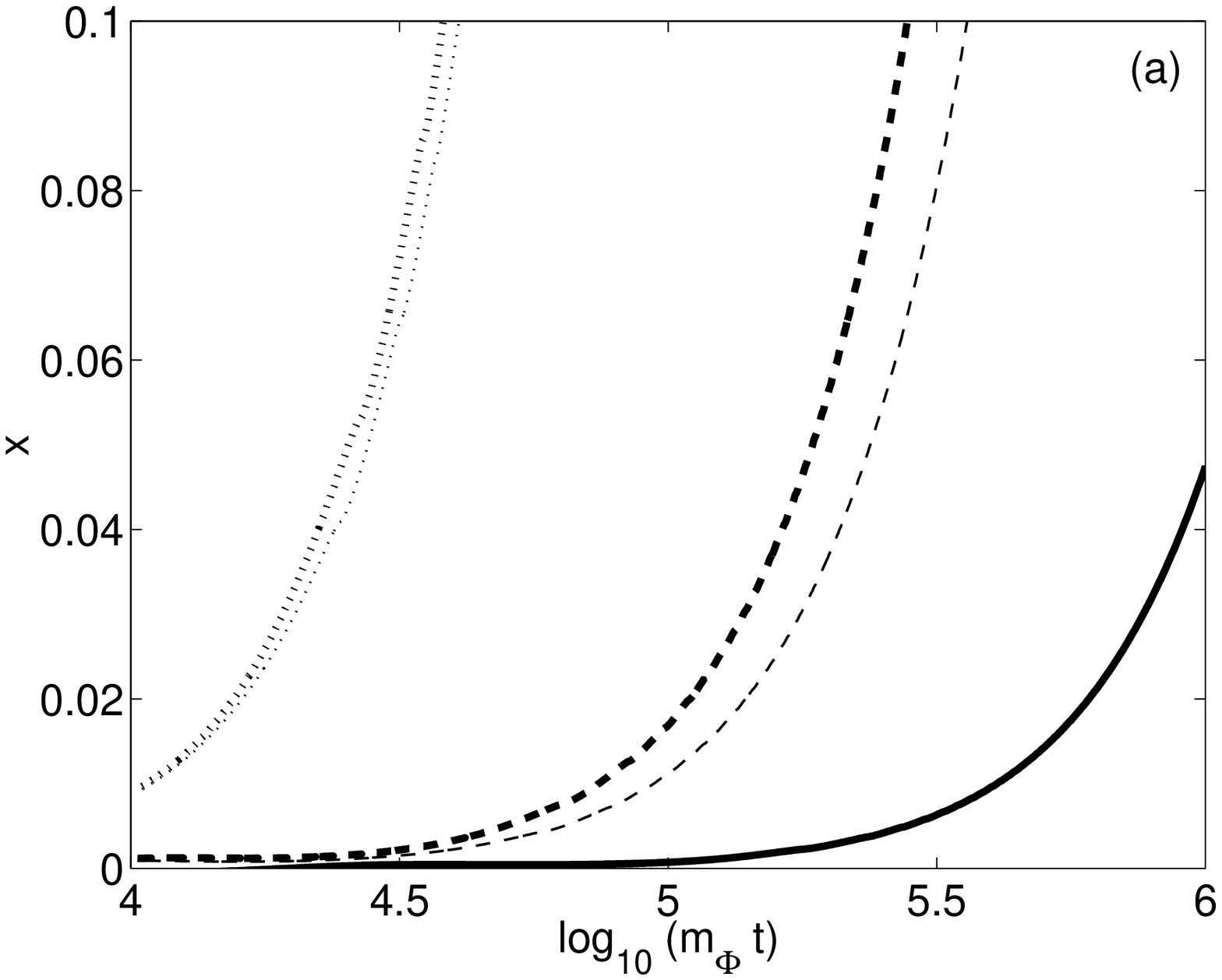}
\includegraphics{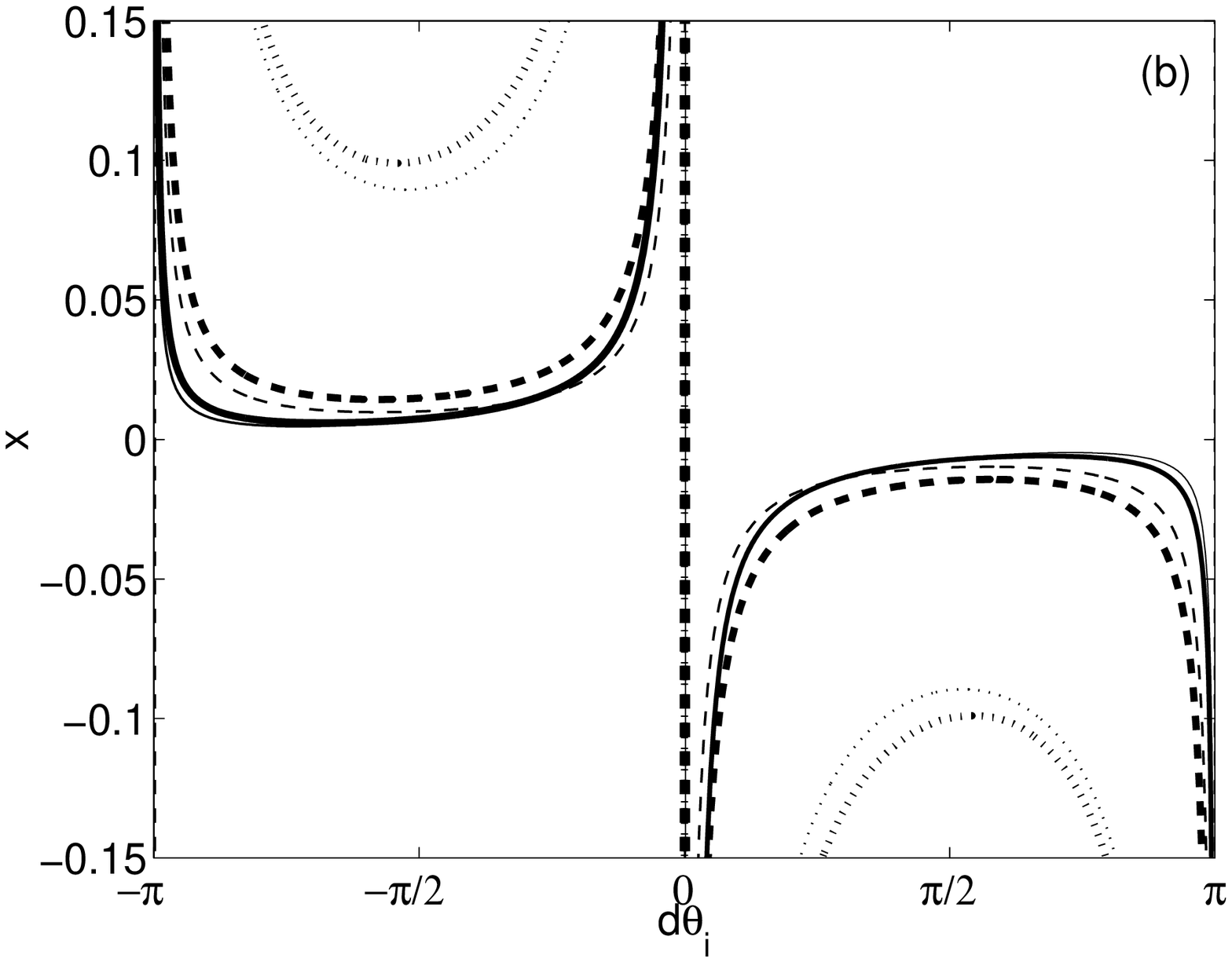}
\includegraphics{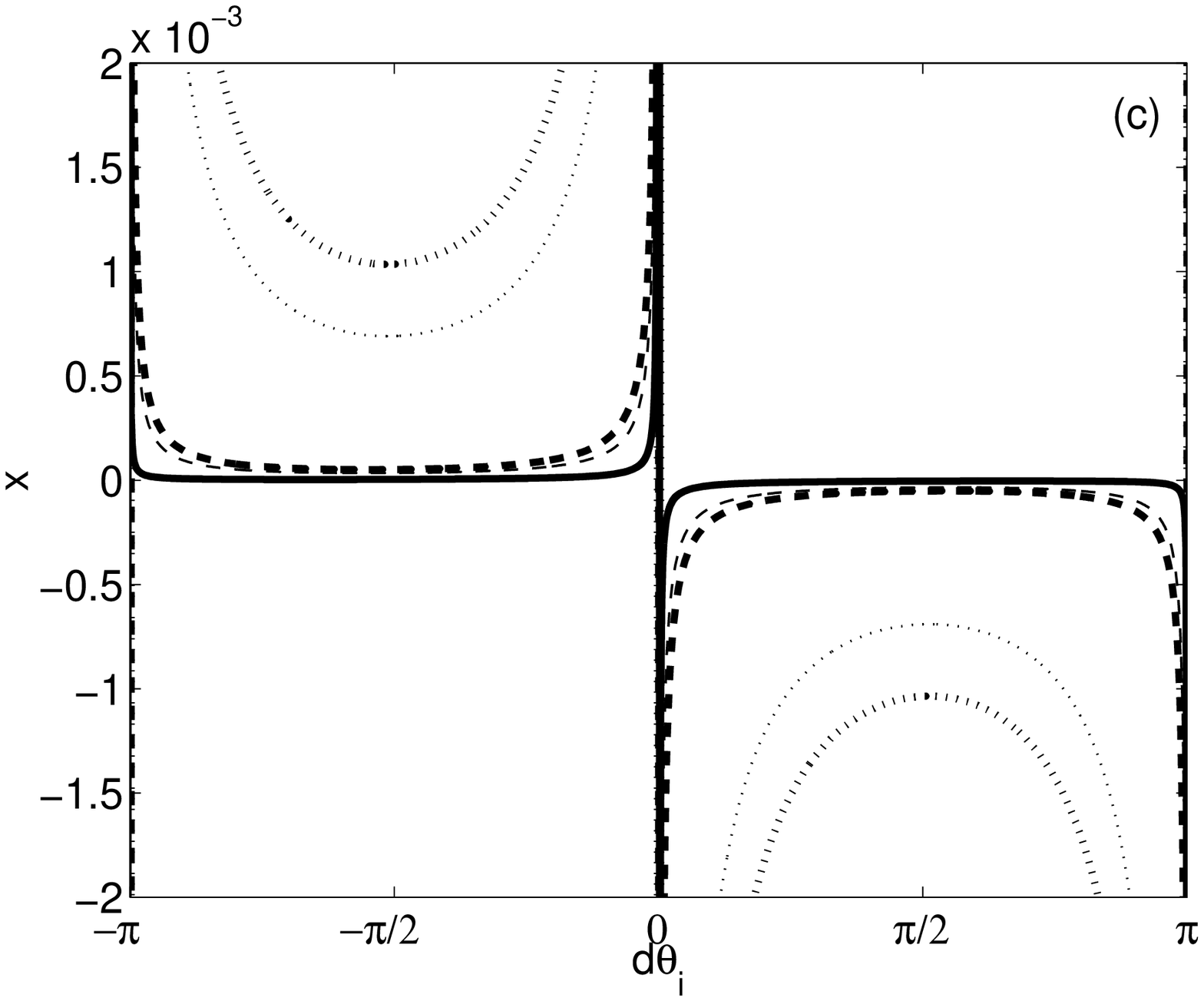}
\caption{\small Energy-to-charge ratio, $x$, in the gauge mediated case vs. (a) time in logarithmic units of time $d=4$, (b) $d=4$ and (c) $d=6$ with D-term (thin lines, $a=0$) and F-term (thick lines, $a=1$) and $m_{\Phi}=1,\,10,\,100\TeV$ (solid, dashed, dotted lines) at $t=4\cdot 10^5m_{\Phi}^{-1},\,10^5m_{\Phi}^{-1},\,4\cdot 10^4m_{\Phi}^{-1}$ ($d=4$) and $t=4\cdot 10^9,\,10^9,\,4\cdot 10^8m_{\Phi}^{-1}$ ($d=6$).}\label{plotenchaga}
\efig

In Figs. \ref{plotpreenga}(a) and (b) we show the time development of the pressure-to-energy density ratio, $w$, for $d=4$ and $d=6$. One can see that the pressure is always negative. The calculation of average pressure is even more involved than in the gravity mediated case, since the oscillation frequency becomes very large. In \fig{plotpreenga}(c) we show the ellipticity as a function of different initial conditions. We note that quite generically $\eps\lsim 0.1$.

\bfig
\leavevmode
\centering
\vspace*{4cm}
\includegraphics{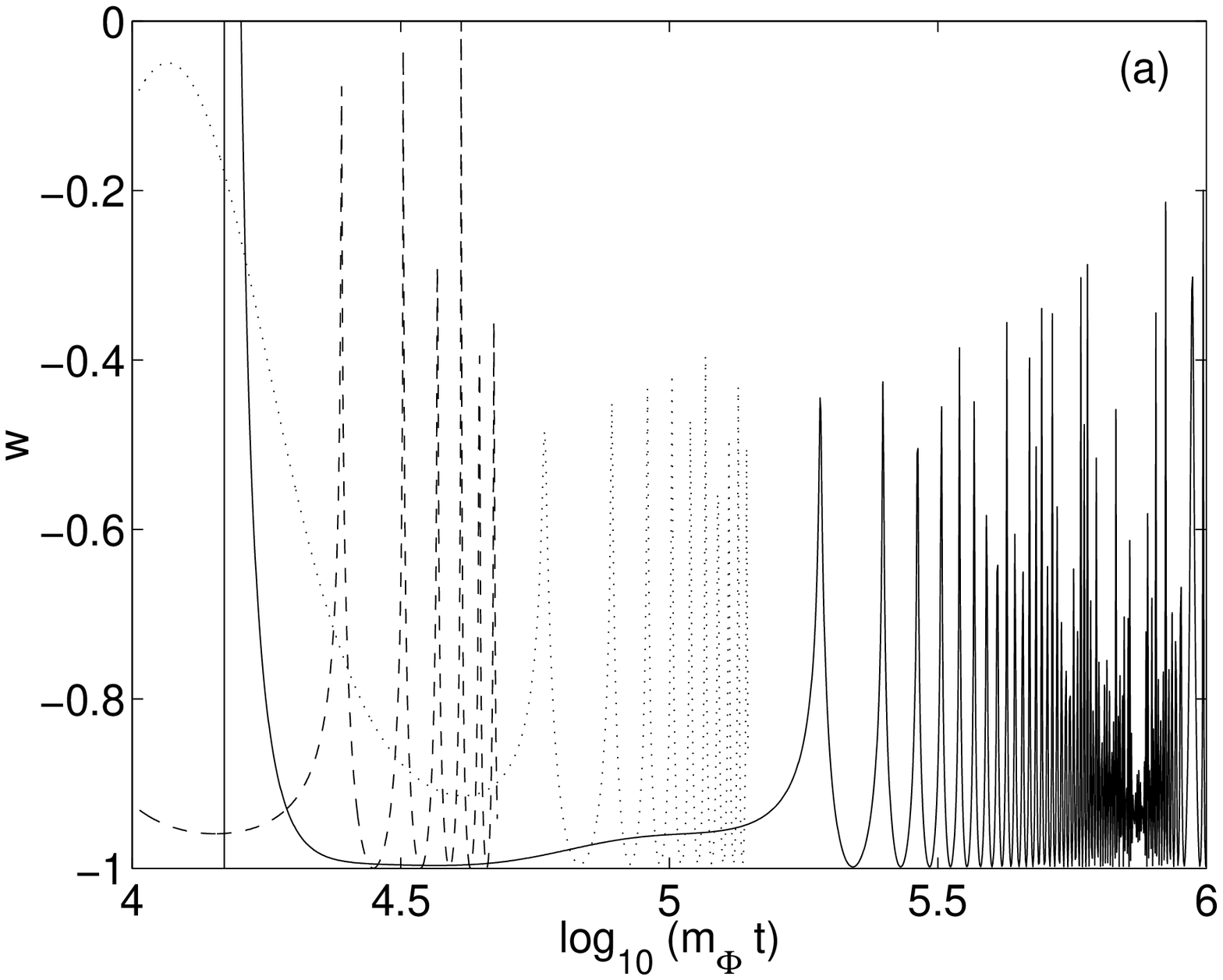}
\includegraphics{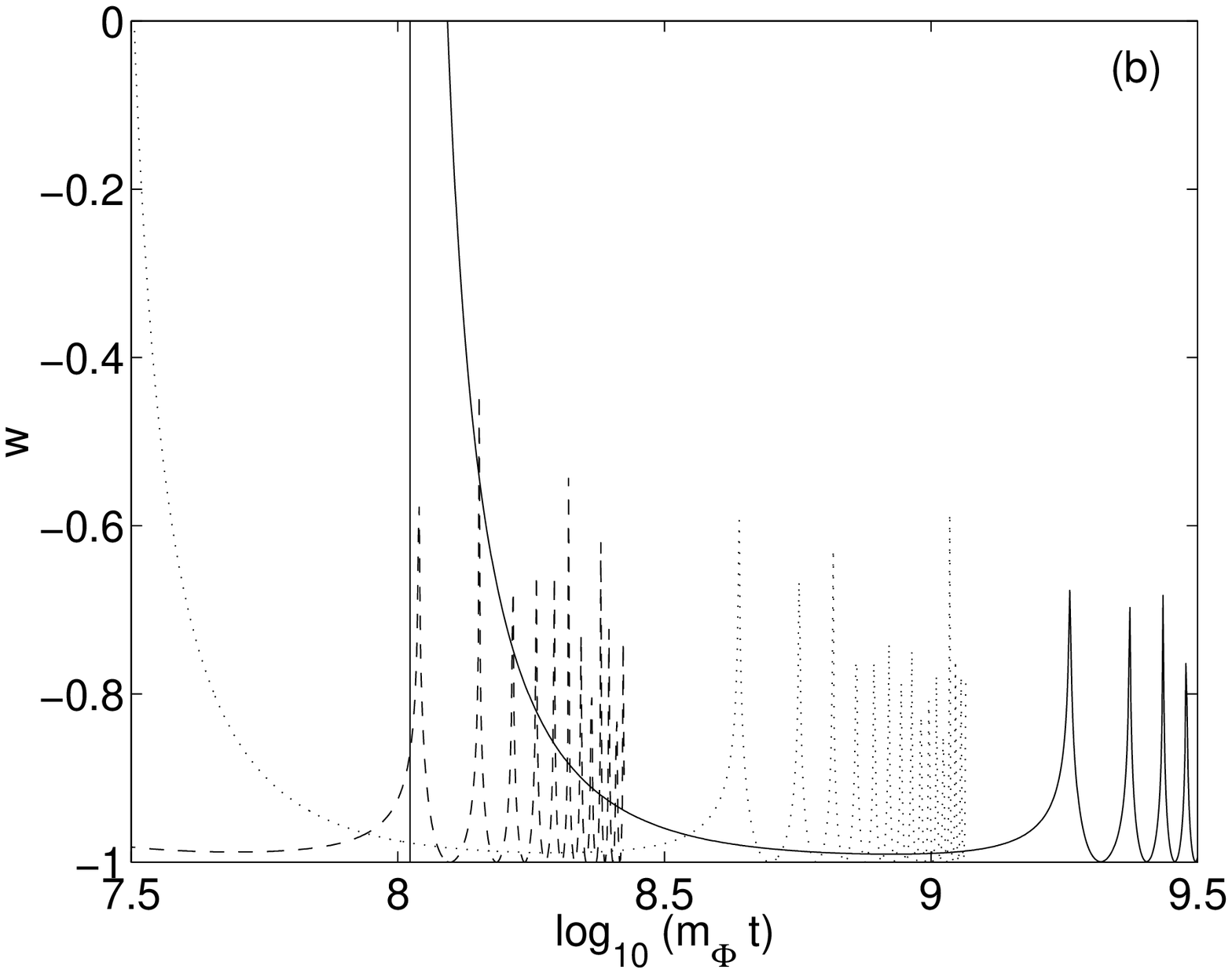}
\includegraphics{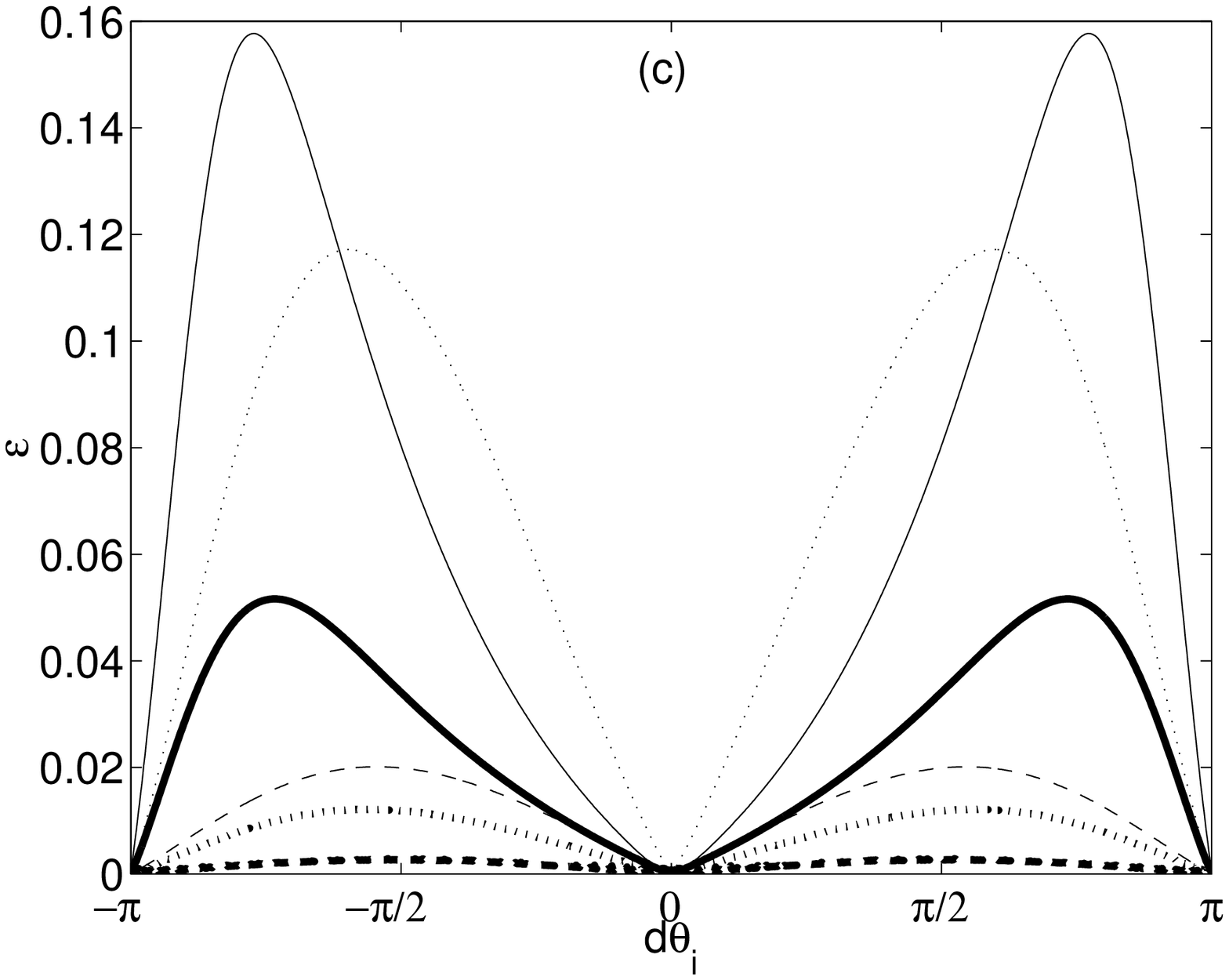}
\caption{\small Pressure-to-energy density ratio, $w$, in the gauge mediated D-term case ($a=0$) vs. time in logarithmic units for (a) $d=4$ and (b) $d=6$. (c) plot of ellipicity of the orbit where $d=4$ (thin lines) and $d=6$ (thick lines). The values of the scalar mass where $m_{\Phi}=1,\,10,\,100\TeV$ with solid, dotted and dashed lines.}\label{plotpreenga}
\efig

In Figs. \ref{plotphigr}(b) and (c) we display the time development of the field amplitude, $\phi(R/R_0)^3$, for $d=4$ and $d=6$. Note that $\phi$ is oscillating, thus confirming the $R^{-3}$ behaviour derived analytically in \eq{fix}.

\subsubsection{Gauge mediated case, F-term}
Now $|a|=1$. This case is essentially the same as the D-term case, as can be gathered from Figs. \ref{plotchaga}, \ref{plotenchaga} and \ref{plotpreenga}. Note again that $|x|$ is slightly larger compared to the D-term case while the orbits are more elliptical, as in the gravity mediated case.

\bc {\bf Table 2. Gauge mediated case} \ec
\btab{|c|c|c|}
\hline
$d$ & $4$ & $6$ \\
\hline
$q(R/R_0)^3\,/\,m_{\Phi}(M_p^{d-3}m_{\Phi}/|\lam|)^{2/(d-2)}$ & $\sim 0.1\ldots 0.6$ & $\sim 1000\ldots 6000$ \\
\hline
$|x|$ & $>10^{-2}$ & $>10^{-5}$ \\
\hline
$w$ & $\sim -0.9\ldots -0.8$ & $\sim -0.95\ldots -0.9$ \\
\hline
$\eps$ & $<0.15$ & $<0.05$ \\
\hline
\etab

\section{Conclusions}
We have studied the evolution of the flat direction field starting from the end of inflation up to the formation of a coherently rotating Affleck-Dine condensate. The original studies \cite{ad,drt} used a mass term $V_m(\Phi)=m^2|\Phi|^2$. We have extended this to radiatively corrected mass terms both in the gravity and gauge mediated case. Typically in both of these cases the mass term grows slower than $|\Phi|^2$. This is the reason why the pressure of the condensate is negative, thus making it unstable with respect to spatial perturbations. The other necessary requirement for a negative pressure was found to be the ellipticity of the rotation of the AD field: for circular orbits the pressure of a coherently rotating field is zero regardless of the potential. Negative pressure is largest for simply oscillating field. Its charge would be zero, resulting in an infinite energy-to-charge ratio, which would make the oscillating configuration the most unstable one. This can also be seen from the analysis of Kasuya and Kawasaki \cite{kskm3} for the growth of perturbation modes: for orbits close to the circular there is only one band of growing modes, but for an oscillating one there appear several growing modes.

We studied numerically the time evolution of various quantities, such as charge density $q$, energy-to-charge ratio $x$, pressure-to-energy density ratio $w$, ellipticity of the orbit $\eps$ and the evolution of the AD field amplitude $\phi$. As an initial condition the homogeneous field was chosen to be at rest at the ground state of the potential. Only the initial phase of the field was varied over all the possible values for both D-term and F-term inflation. We found that there is no qualitative difference between the two cases. We checked both the gravity mediated case and the gauge mediated case with the dimension of the non-renormalizable operators $d=4,\,5,\,6,\,7$. (For $d=4$, $K=0$ our results agree with \cite{drt}.) In contrast to \cite{drt} the energy-to-charge ratio, $x$, was found not to asymptote to a constant value. Instead $|x|$ follows an increasing curve due to negative pressure with a minimum $|x|>1.1$ in the gravity mediated case, $|x|>10^{-2}$ for $d=4$ and $|x|>10^{-5}$ for $d=6$ in the gauge mediated case. We should mention that the calculation of the average of $w$ is plagued with technical difficulties. However, the qualitative behaviour of $w$ is not dependent on these. We believe we have managed to check numerically that the ellipticity of the orbit really affects the pressure exactly the way indicated by the analytical considerations. In addition the field amplitude $\phi$ was shown to decay $R^{-6/(n+2)}$ for potentials $V(\phi)\sim \phi^n$, where $n=0$ would correspond to the gauge mediated case, which is in exact analogy to the coherent oscillation \cite{turner}.

In conclusion, we have shown that the AD condensate is unstable due to negative pressure, which is caused by two effects: the potential growing slower than the field squared and the ellipticity of the orbit. We have also shown that after the fragmentation of the AD condensate the thermalization of the resulting Q-ball distribution depends on the energy-to-charge ratio of the condensate. In the gravity mediated case $x$ was found to be  such that thermalization is likely to be a generic feature, as suggested by analytical considerations. The gauge mediated case is complicated by the fact that condensate formation can take place while the field is located in the part of the potential dominated by gravity mediated mass term, whereas Q-balls can be created by virtue of gauge mediation. However, as discussed Sect. 3.2.3, thermalization appears to be very likely also in this case.

\subsection*{Acknowledgements}
The author would like to thank K. Enqvist for discussions. This work has been partly supported by the Academy of Finland under the contract 101-35224.

\end{document}